\documentclass[letterpaper,11pt]{article}
\usepackage{jheppubMod}
\usepackage[english]{babel}
\usepackage{amsmath,amssymb,graphicx,float,slashed,xcolor,multicol}
\usepackage{graphicx,tabularx}
\usepackage{url}
\usepackage{footmisc}
\usepackage{amsfonts}
\usepackage{cancel}
\usepackage{color}
\usepackage{multirow} 
\usepackage{pifont}
\usepackage{epstopdf}
\usepackage{slashed}
\usepackage{comment}
\usepackage{booktabs} 
\usepackage{array}
\usepackage{mathrsfs}
\usepackage{subfigure}
\usepackage[toc,page]{appendix}
\usepackage{mathtools}
\usepackage{romannum}
\usepackage{ulem}
\usepackage{bbold}
\usepackage{enumitem}
\usepackage{multirow}
\usepackage{amsmath}
\usepackage{graphicx}
\usepackage{lineno}
\graphicspath{{Plots/}}

\hypersetup{bookmarks=true,unicode=true,pdftoolbar=true,pdfmenubar=true,
  pdffitwindow=false,pdfstartview={FitH},
  pdftitle={BSM $\nu$ physics: complementarity across energies},pdfauthor={Han, et al},
  pdfsubject={Subject},pdfcreator={Creator},pdfproducer={Producer},
  pdfkeywords={Neutrino masses}{NSI}{Seesaw Models}{BSM},
  pdfnewwindow=true,colorlinks=true
}

\def\beq{\begin{equation}}
\def\eeq{\end{equation}}
\def\lag{\mathscr{L}}

\title{BSM $\nu$ physics: complementarity across energies  \\ 
$-$ a white paper for Snowmass 2021}

\author[a]{Tao Han,}
\author[b]{Jiajun Liao,}
\author[c]{Hongkai Liu,}
\author[d,e]{Danny Marfatia,}
\author[f]{Richard Ruiz}

\affiliation[a]{PITT PACC, Department of Physics and Astronomy,\\ University of Pittsburgh, Pittsburgh, PA 15260, USA }
\affiliation[b]{School of Physics, Sun Yat-Sen University, Guangzhou, 510275, China}
\affiliation[c]{Department of Physics, Technion – Israel Institute of Technology, Haifa 3200003, Israel}
\affiliation[d]{Department of Physics and Astronomy, University of Hawaii at Manoa, Honolulu, HI 96822, USA}
\affiliation[e]{Kavli Institute for Theoretical Physics, University of California, Santa Barbara, CA 93106, USA}
\affiliation[f]{Institute of Nuclear Physics, Polish Academy of Sciences (IFJ PAN),\\ ul. Radzikowskiego, Krak{\'o}w 31-342, Poland}
\emailAdd{than@pitt.edu}
\emailAdd{liaojiajun@mail.sysu.edu.cn}
\emailAdd{hol42@pitt.edu}
\emailAdd{dmarf8@hawaii.edu}
\emailAdd{rruiz@ifj.edu.pl}

\preprint{IFJPAN-IV-2022-3, PITT-PACC-2203}

\abstract{We reiterate that there is significant complementarity between low-energy experiments and high-energy colliders in exploring new physics associated with neutrino properties and their mass generation mechanisms. Signals of the new physics in the two energy regimes may be correlated with each other from the same underlying dynamics. We demonstrate the complementary nature by presenting the physics reaches for the Seesaw models of Type I, II and III, and for general neutrino interactions in an effective field theory framework, and in a $Z'$ model.
}

\begin{document}
\maketitle
%%%%%%%%%%%%%%%%%%%%%%%%%%%%%%%%%%%%%%%%%%%%%%%%%%%%%

\section{Introduction}

Flavor oscillations between massive neutrinos is a firmly established phenomenon that cannot
be accounted for by the Standard Model (SM). Neutrino oscillations thus strongly motivate physics beyond
the SM (BSM) that is associated with the neutrino sector. New physics associated with neutrinos can be probed at different energy scales and with limited theoretical prejudice.
Certain signals of the new physics in the two energy regimes may be correlated with each other from the same underlying theory. 
It is ultimately important to seek complementary signals to establish a consistent picture of the underlying physics.

There are three tree-level mechanisms to generate neutrino masses of Majorana type that extend the SM's field content in a minimal manner~\cite{Ma:1998dn}.
All of these introduce new particle states, including ones with gauge charges, and thus lead to rich phenomenology, complementary in different experiments at all energy scales. Here, we summarize 
searches for testing these models at the Large Hadron Collider (LHC) and its upgrades. For  wide-ranging reviews on tests of both tree-level and radiative neutrino mass models, see Refs.~\cite{Deppisch:2015qwa,Cai:2017jrq,Cai:2017mow}.

On the phenomenological ground, general neutrino interactions (GNI) is the most general parameterization of new physics in the neutrino sector and widely used at the low-energy probes.
The effective four-fermion interactions are commonly written as
\beq
\lag_{\text{GNI}}= \frac{G_F}{\sqrt{2}}\sum_a (\bar\nu\Gamma_a\nu )[\bar f\Gamma_a(\epsilon_a + \Tilde{\epsilon}_a\gamma^5) f],
\eeq
where $\Gamma_a = \{\mathbb{1}, i\gamma^5, \gamma^\mu, \gamma^\mu\gamma^5, \sigma^{\mu\nu}\}$ are the five Lorentz structures, corresponding to scalar, pseudoscalar, vector, axialvector, and tensor interactions, respectively. 
$f$ denotes quarks and charged leptons. 
$\epsilon$ and $\tilde{\epsilon}$ are the effective GNI parameters, and contain four flavor indices, which are kept implicit. 
The GNI is not valid above the weak scale or new physics scale. For illustration of the UV completion, we introduce a neutral gauge boson $Z'$ to show the effects at different energies. We then adopt an SM effective field theory formalism including a right-handed neutrino ($\nu$SMEFT/SMNEFT). We demonstrate that the low- and high-energy probes have their own merits based on Refs.~\cite{Han:2019zkz,Han:2020pff,Cirigliano:2021peb}.

%%%%%%%%%%%%%%%%%%%%%%%%%%%%%%%%%%%%%%%%%%%%%%%%%%%%%%%%%%%%%%
\section{Collider tests of tree-level seesaws}

Collider experiments offer powerful probes of heavy states predicted by neutrino mass models. As benchmarks, direct tests of minimal tree-level scenarios typically focus on phenomenological implementations~\cite{delAguila:2008cj,Atre:2009rg} of the canonical Types I, II, and III Seesaw models. Phenomenological scenarios treat the masses and mixing of new particles states as independent parameters to minimize model dependence. 

%%%%%%%%%%%%%%%%%%%%%%%%%%%%%%%%%%%%%%%%%%%%%%%%%%%%%%%%%%%%%%
{\bf The Phenomenological Type I Seesaw} extends the SM field content by $n_R\geq2$ right-handed neutrinos $\nu_R$. After electroweak (EW) symmetry breaking, the model parameterizes the mixing between EW currents and neutrino mass eigenstates by the decomposition
\begin{equation}
    \nu_\ell = \sum_{m=1}^3 U_{\ell m}^* \nu_m + \sum_{m'=4}^{3+n_R} V_{\ell m'}^* N_{m'},
\end{equation}
where $\nu_\ell$ is a SM neutrino of flavor $\ell$ in the interaction basis, $\nu_m$ are the three light neutrino mass eigenstates, $U_{\ell m}$ are the light-active mixing matrix elements measured in oscillation experiments, $N_{m'}$ are heavy mass eigenstates, and $V_{\ell m'}$ are active-heavy (or active-sterile) mixing matrix elements. For discovery purposes, one typically considered only the $N_{m'=4}$ states and decouples the remaining $(n_R-1)$ heavy mass eigenstates. The states $N_k$ couple to EW bosons through mixing, and therefore can be produced in a variety of mechanisms.
At the LHC, the leading channels are the Drell-Yan mechanism~\cite{Keung:1983uu},
$W\gamma$ fusion~\cite{Datta:1993nm,Dev:2013wba,Alva:2014gxa,Degrande:2016aje},
gluon fusion~\cite{Dicus:1991wj,Hessler:2014ssa,Ruiz:2017yyf},
and
same-sign $W^\pm W^\pm$ scattering~\cite{Dicus:1991fk,Fuks:2020att}.
For recent comparisons of the channels at the LHC and beyond, see Refs.~\cite{Degrande:2016aje,Pascoli:2018heg,Fuks:2020att}.

\begin{figure}[t!]
\begin{center}
\subfigure[]{\includegraphics[width=0.48\textwidth]{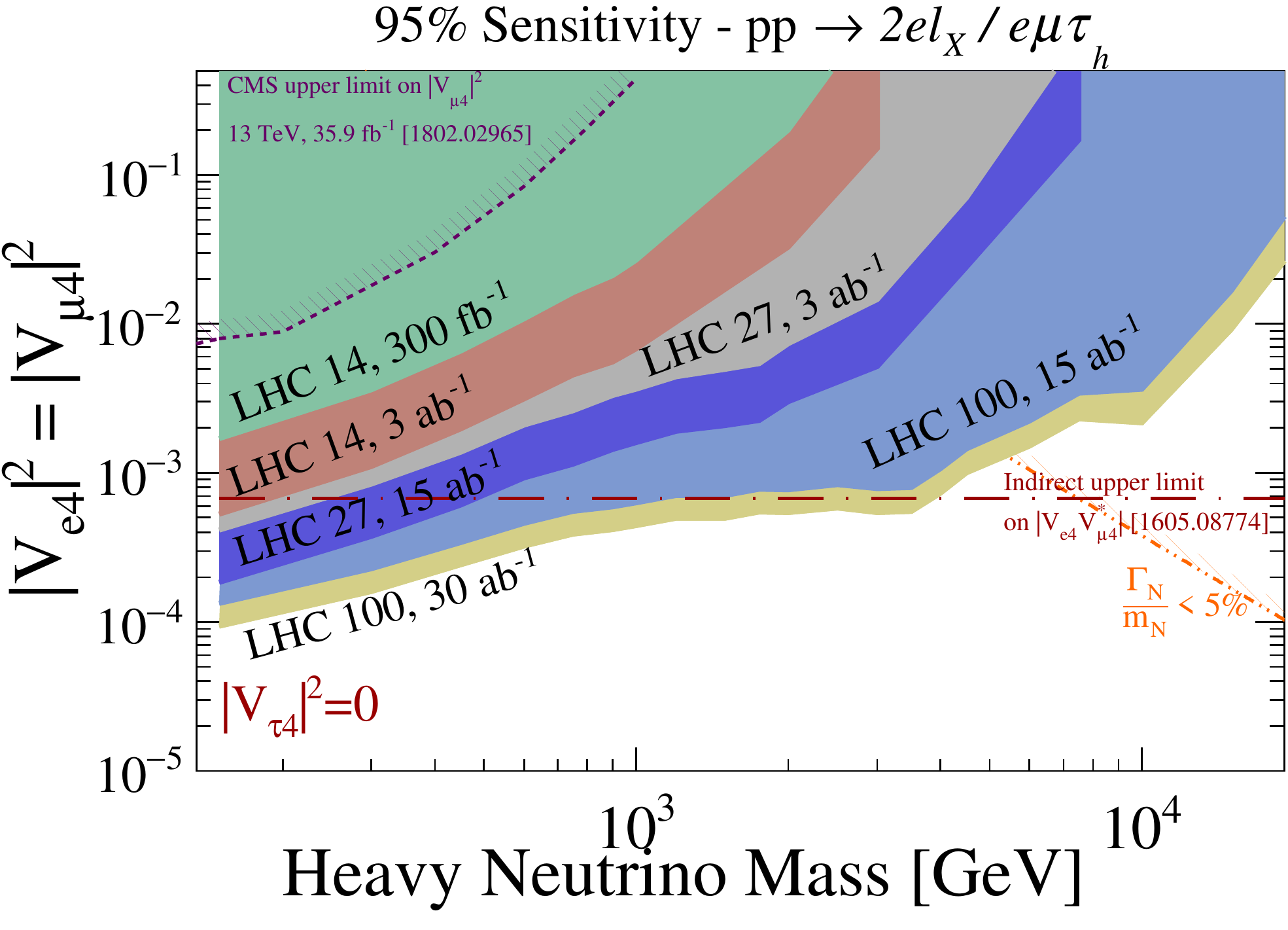}\label{fig:outlook_typeI_trilepton_muel}}
\subfigure[]{\includegraphics[width=0.48\textwidth]{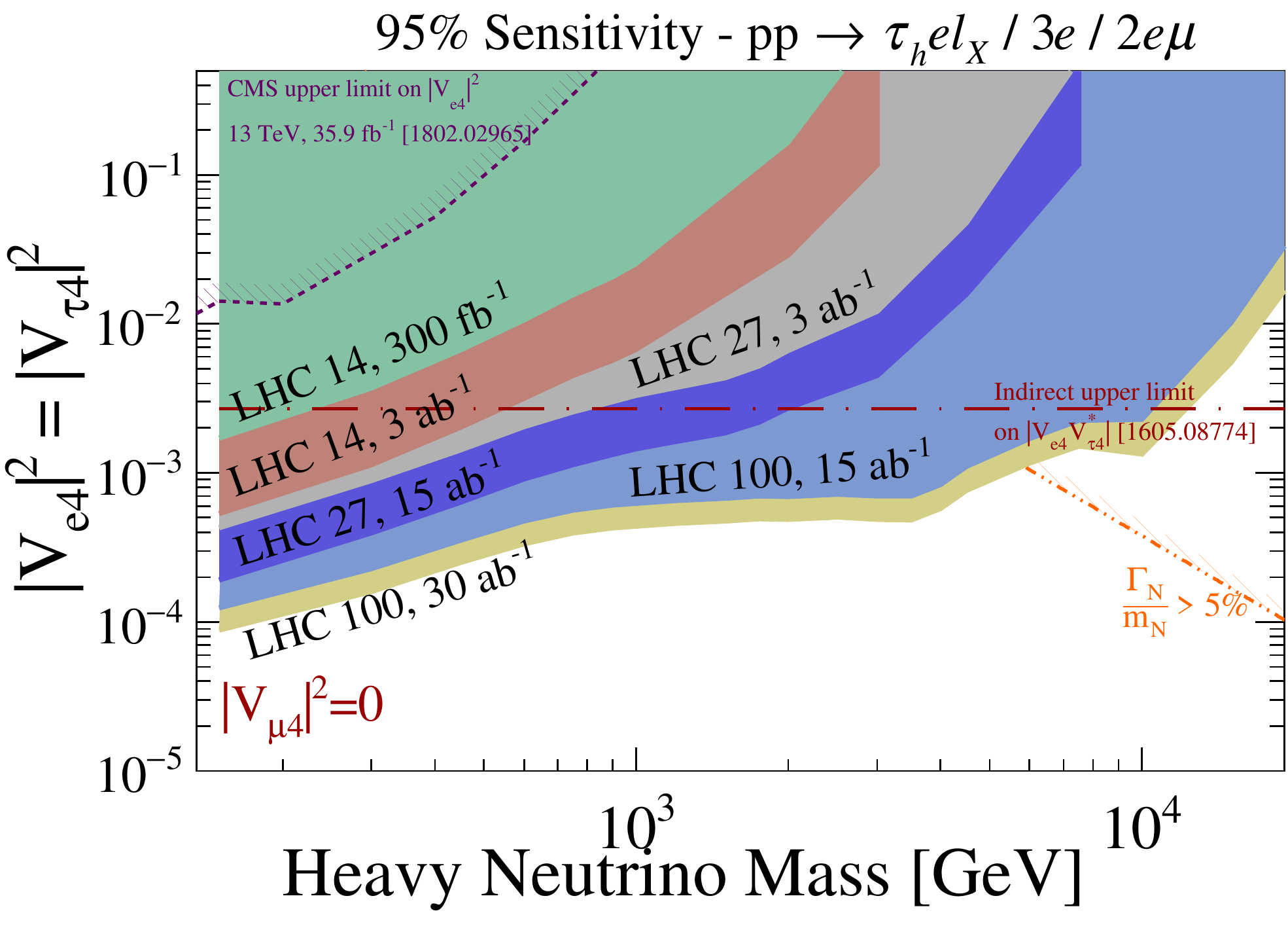}\label{fig:outlook_typeI_trilepton_tael}}
\\
\subfigure[]{\includegraphics[width=0.48\textwidth]{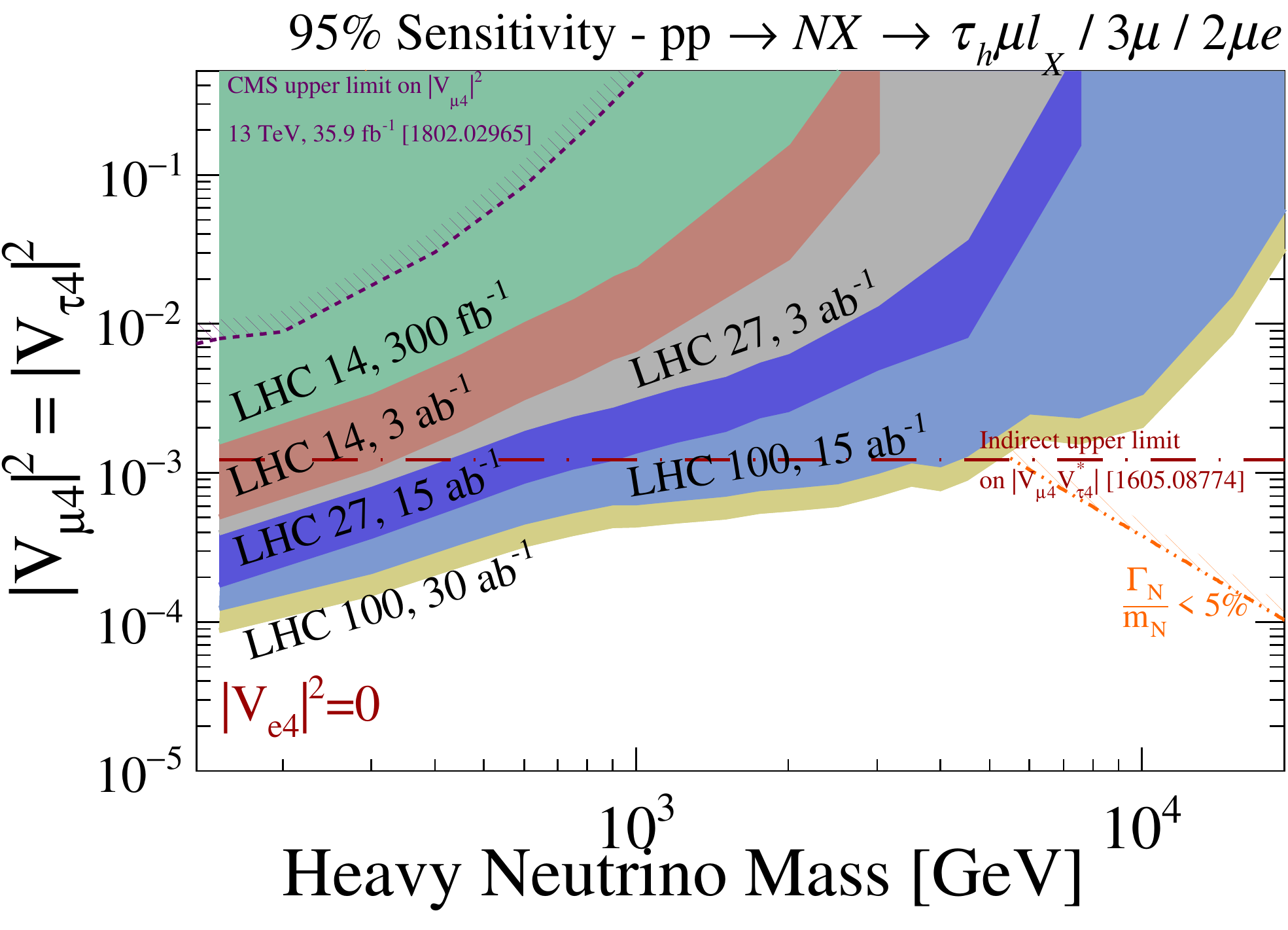}\label{fig:outlook_typeI_trilepton_tamu}}
\subfigure[]{\includegraphics[width=0.48\textwidth]{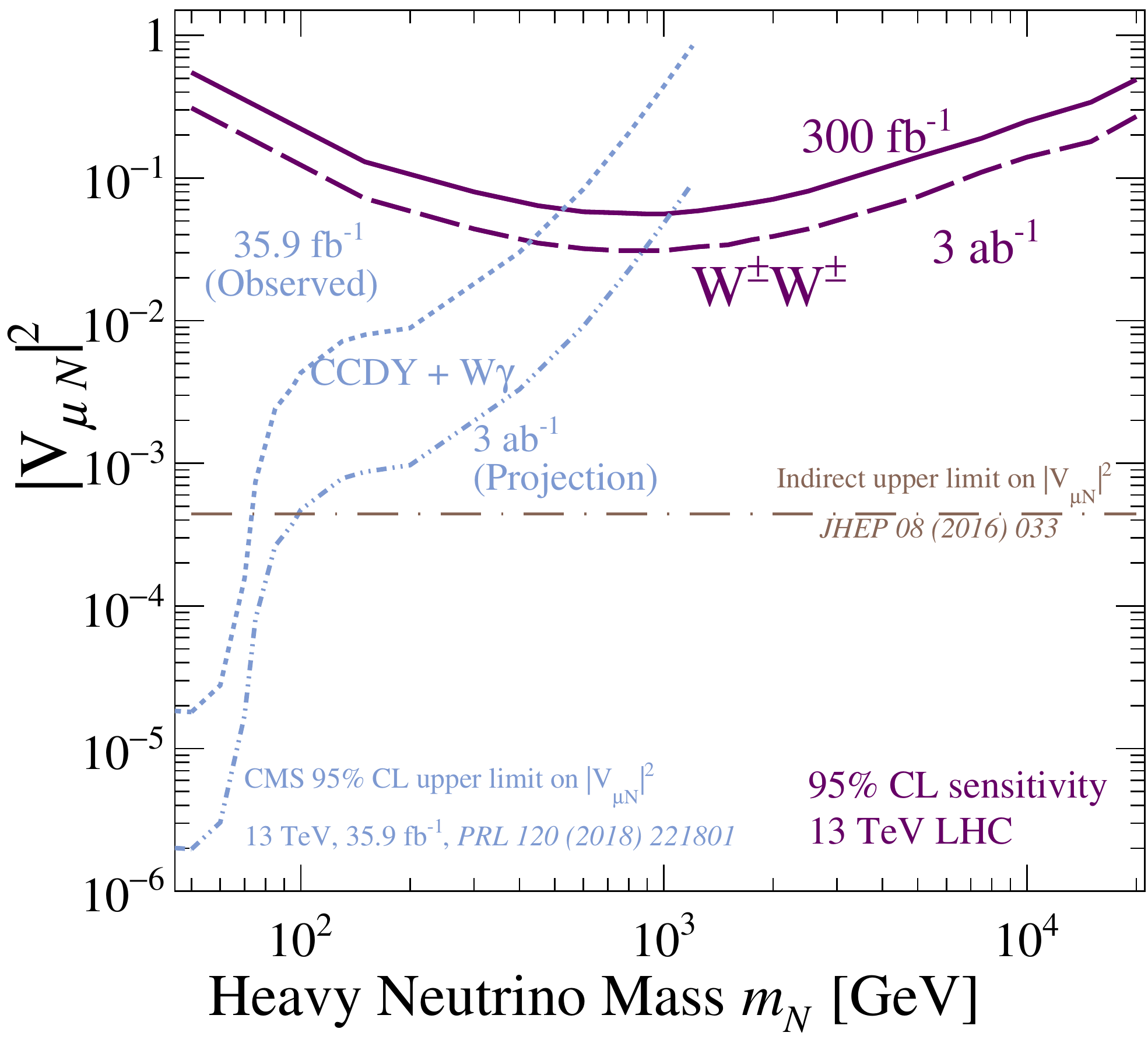}\label{fig:outlook_typeI_WWdilepton}}
\end{center}
	\caption{95\% CL sensitivity to active-sterile mixing $\vert V_{\ell 4}\vert^2$ as a function heavy neutrino mass at the LHC in the trilepton and MET channel using the analysis of Refs.~\cite{Pascoli:2018rsg,Pascoli:2018heg} for the benchmark flavor categories: 
	(a) $\vert V_{e4}\vert^2 = \vert V_{\mu 4}\vert^2$ with $\vert V_{\tau 4}\vert^2=0$,
	(b) $\vert V_{e4}\vert^2 = \vert V_{\tau 4}\vert^2$ with $\vert V_{\mu 4}\vert^2=0$,
	(c) $\vert V_{\mu4}\vert^2 = \vert V_{\tau 4}\vert^2$ with $\vert V_{e 4}\vert^2=0$.
	(d) Same but using the same-sign $W^\pm W^\pm\to\mu^\pm\mu^\pm$ scattering channel with the analysis of Ref.~\cite{Fuks:2020att}.
	}
	\label{fig:outlook_typeI}
\end{figure}

Figure~\ref{fig:outlook_typeI} shows the 95\% CL sensitivity to active-sterile mixing $\vert V_{\ell 4}\vert^2$ as a function heavy neutrino mass at the LHC in the trilepton and MET channel using the analysis of Refs.~\cite{Pascoli:2018rsg,Pascoli:2018heg} for the benchmark flavor categories: 
	(a) $\vert V_{e4}\vert^2 = \vert V_{\mu 4}\vert^2$ with $\vert V_{\tau 4}\vert^2=0$,
	(b) $\vert V_{e4}\vert^2 = \vert V_{\tau 4}\vert^2$ with $\vert V_{\mu 4}\vert^2=0$,
	(c) $\vert V_{\mu4}\vert^2 = \vert V_{\tau 4}\vert^2$ with $\vert V_{e 4}\vert^2=0$.
Depending on the precise flavor category, active-sterile mixing as small as  $\vert V_{\ell N}\vert^{2} \sim 10^{-4}-10^{-2}$ can be reached for masses in the approximate range of $m_N \approx 200-1200$ GeV with 3 ab$^{-1}$ of data~\cite{Pascoli:2018rsg,Pascoli:2018heg}.
For some signal categories, the trilepton becomes directly competitive with low-energy tests of charged lepton flavor violation~\cite{Fernandez-Martinez:2016lgt}.
Fig.~\ref{fig:outlook_typeI}(d) shows the same but using the same-sign $W^\pm W^\pm\to\mu^\pm\mu^\pm$ scattering channel with the analysis of Ref.~\cite{Fuks:2020att}. 
The analysis shows that mixing as small as  $\vert V_{\ell N}\vert^{2} \sim 10^{-1}$ can be reached for as large as $m_N \sim \mathcal{O}(10)$ TeV. It is important to stress that, if discovered, measuring heavy neutrinos' chiral couplings will be paramount to fully exploring their properties and potential couplings to other undiscovered particles~\cite{Han:2012vk,Ruiz:2017nip,Ruiz:2020cjx}.

%%%%%%%%%%%%%%%%%%%%%%%%%%%%%%%%%%%%%%%%%%%%%%%%%%%%%%%%%%%%%%
{\bf The Type II Seesaw} extends the SM field content by a scalar multiplet that is a triplet under SU$(2)_L$ and carries hypercharge $Y=1$. In the gauge basis, this can be  written as
\begin{equation}
    \sqrt{2}\hat{\Delta} =
\begin{pmatrix}
\hat{\Delta}^+ & \sqrt{2} \hat{\Delta}^{++} \\ 
\sqrt{2} \hat{\Delta}^0 & -\hat{\Delta}^+ 
\end{pmatrix},
\end{equation}
which couples at tree-level to the SM Higgs field, including through a dimension $d=4$ operator  that explicitly violates lepton number. After EW symmetry breaking, $\hat\Delta$ acquires a vacuum expectation value (vev) and spontaneously generates left-handed Majorana masses for neutrinos through a Yukawa coupling. Importantly, this is an instances of generating neutrino masses without right-handed neutrinos.

\begin{figure}[t!]
\begin{center}
\includegraphics[width=\textwidth]{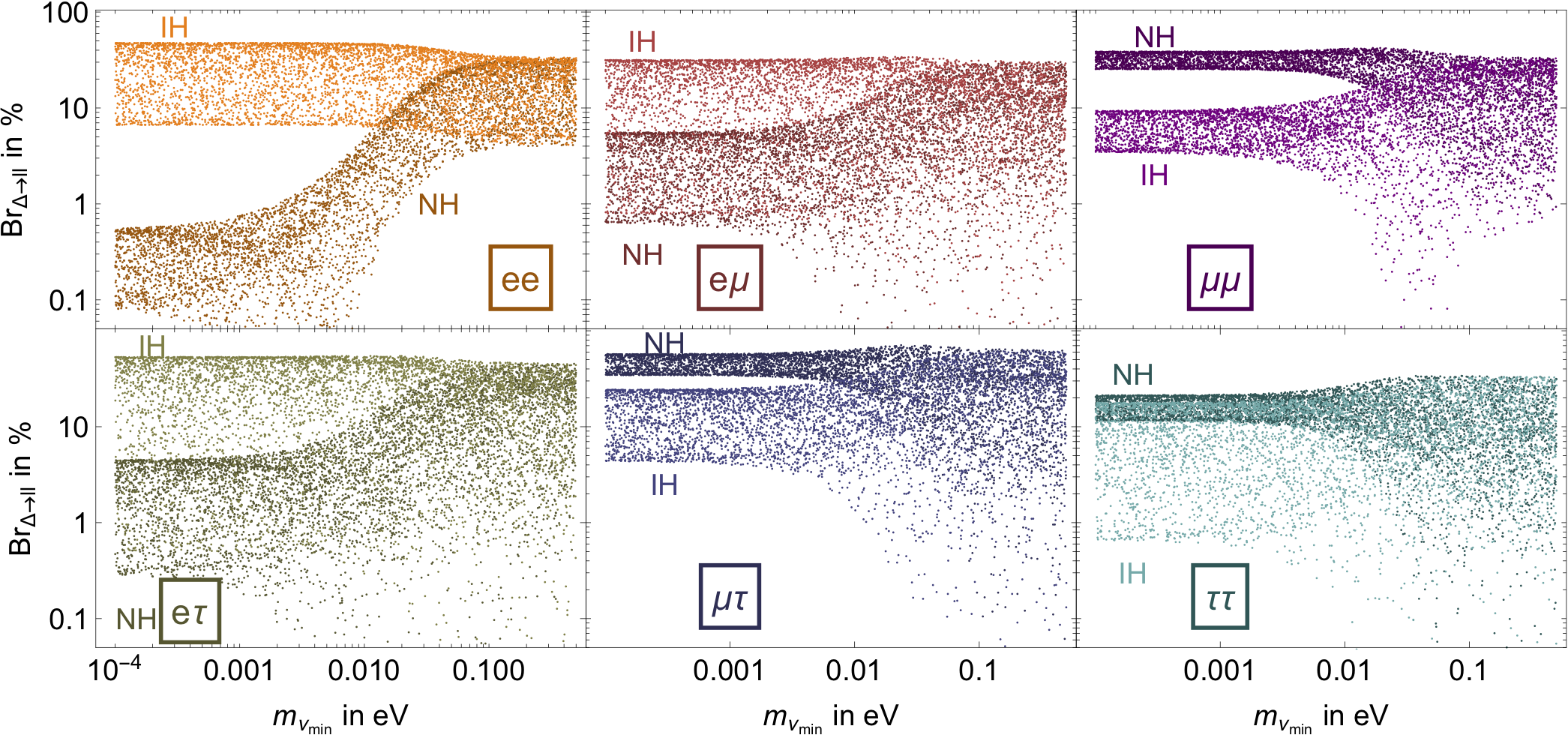}
\end{center}
	\caption{The branching rate (BR) of $\Delta^{\pm\pm}\to \ell^\pm_i\ell^\pm_j$ decays in the Type II Seesaw as a function of the lightest neutrino mass for the normal hierarchy (NH) and inverted hierarchy (IH) ordering of neutrino masses, and in the limit of a vanishing triplet vev. Adapted from Ref.~\cite{Fuks:2019clu}.}
	\label{fig:typeIInlo_BrDpp_All}
\end{figure} 

After mixing with the SM Higgs, the Type II Seesaw predicts the existence of four new scalar, mass eigenstates: a doubly electrically charged Higgs $\Delta^{\pm\pm}$, a singly electrically charged Higgs $\Delta^\pm$, an electrically neutral CP-even Higgs $\Delta^0$, and an electrically neutral CP-odd Higgs $\xi^0$. These new states carry EW gauge couplings and can be produced through  variety of tree- and loop-induced processes~\cite{Nemevsek:2016enw,Fuks:2019clu}. For a recent comparison of these channels at the LHC and future proton colliders, see Ref.~\cite{Fuks:2019clu}.

Due to the absence of sterile neutrinos, the flavor structure and mass pattern of neutrinos are entirely set by the Yukawa couplings between $\hat{\Delta}$ and the leptonic doublets of the SM. This means that the decay rates of $\Delta^{\pm\pm},\ \Delta^{\pm},\ \Delta^0,\ $ and $\xi^0$ to leptons, which are proportional to Yukawa couplings, are strongly constrained by the PMNS matrix, and hence neutrino oscillation data~\cite{FileviezPerez:2008jbu}.
For the case of $\Delta^{\pm\pm}$, Fig.~\ref{fig:typeIInlo_BrDpp_All} shows~\cite{Fuks:2019clu} the branching rate (BR) of $\Delta^{\pm\pm}\to \ell^\pm_i\ell^\pm_j$ decays as a function of the lightest neutrino mass for the normal hierarchy (NH) and inverted hierarchy (IH) ordering of neutrino masses, in the limit of a vanishing triplet vev. The correlations predicted for the decay fractions of Type II scalars highlights the complementarity of oscillation experiments and high-energy collider experiments.

\begin{figure}[t!]
\begin{center}
\subfigure[]{\includegraphics[width=0.48\textwidth]{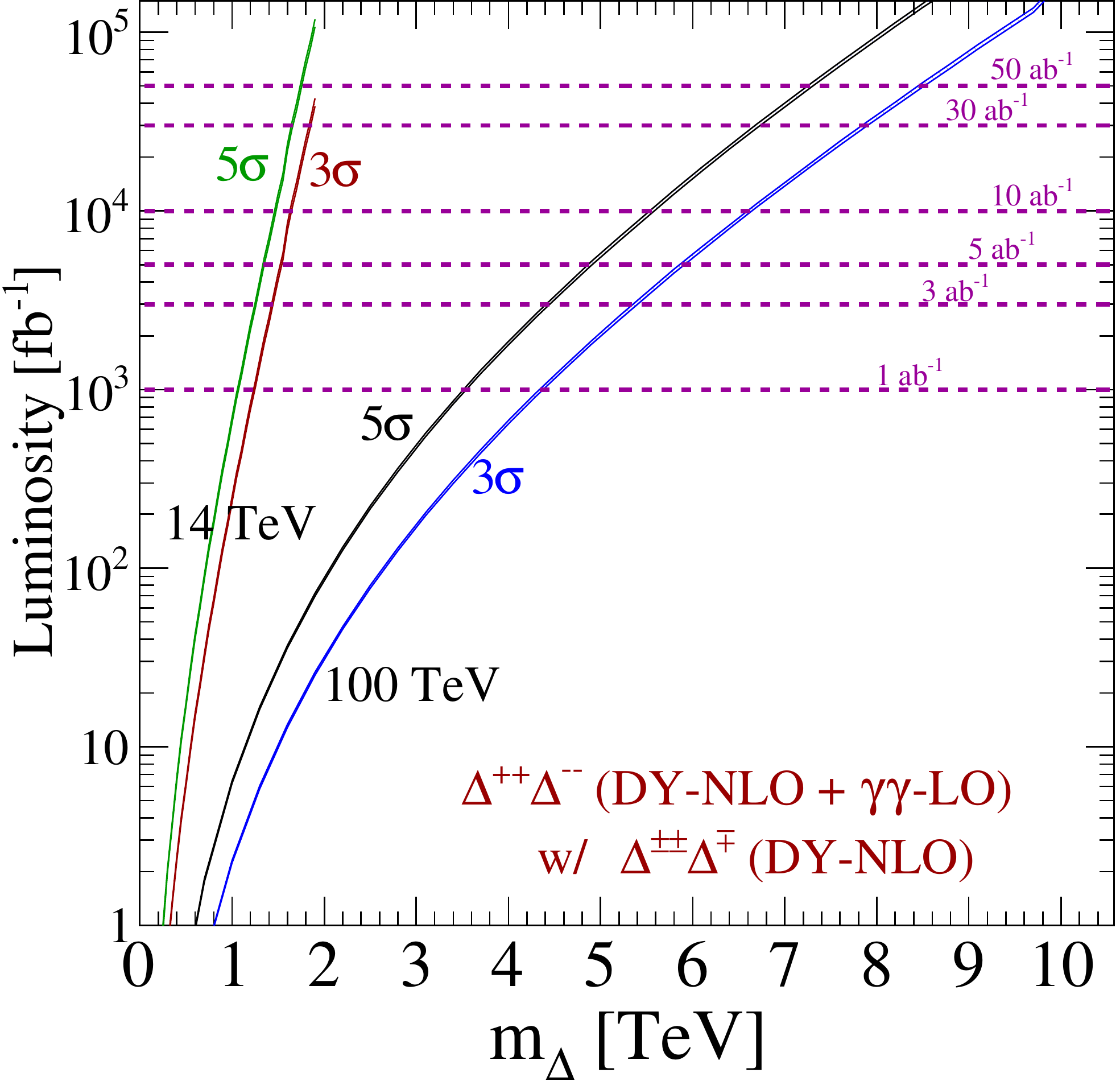}\label{fig:typeIInlo_Lumi_vs_mD_LHCMulti}}
\subfigure[]{\includegraphics[width=0.48\textwidth]{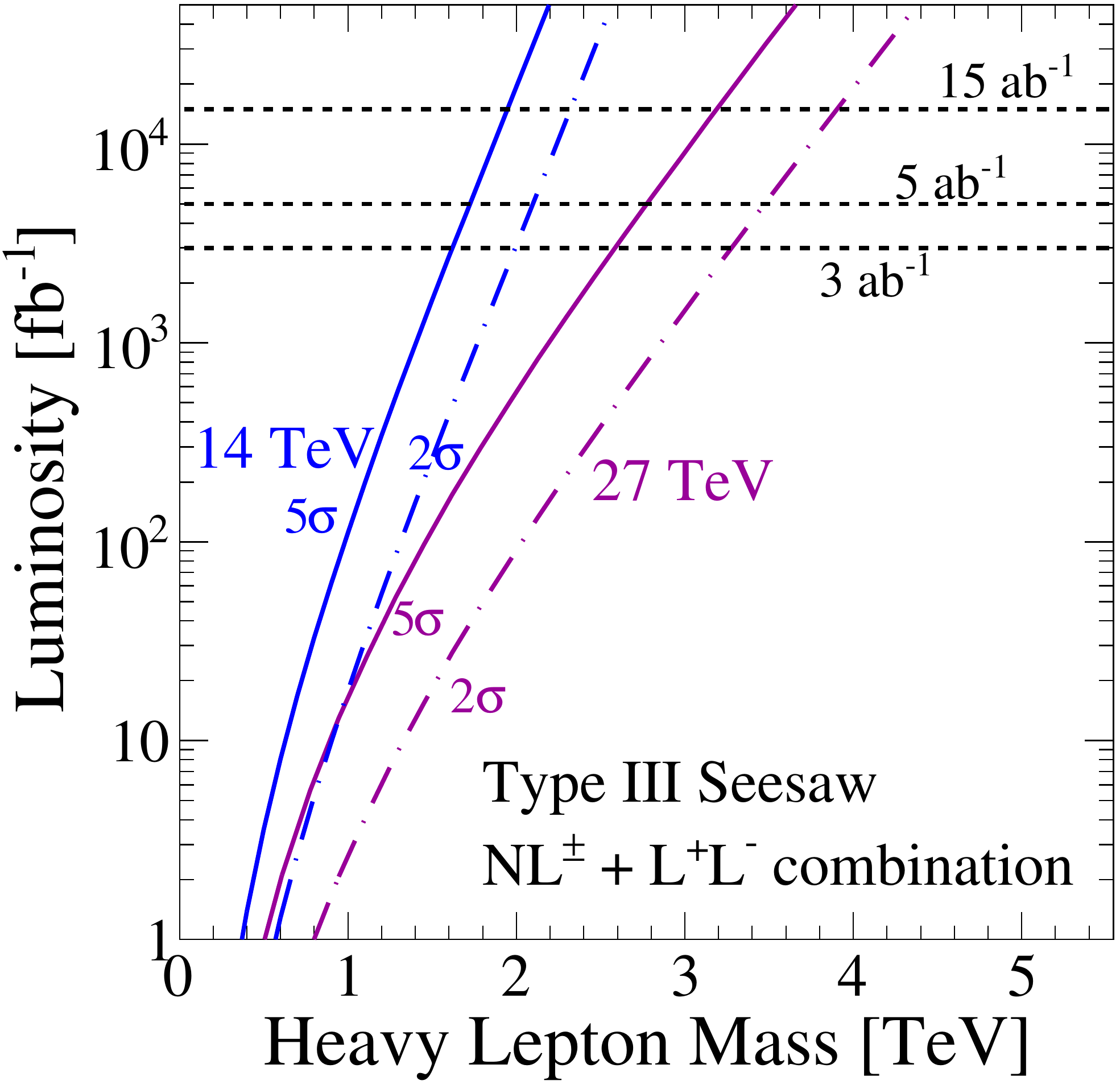}\label{fig:type3_HLvsHL_LHC_Disc_vs_Mass}}
\end{center}
	\caption{
	(a) The  luminosity required to reach $5\sigma~(3\sigma)$ discovery (sensitivity) of Type II Seesaw scalars produced in pairs produced in $\sqrt{s}=14$ and 100 TeV $pp$ collisions, and which subsequently decay to multi-lepton final states. Adapted from Ref.~\cite{Fuks:2019clu}.
	(b) The same but for Type III leptons produced in pairs in $\sqrt{s}=14$ and 27 TeV $pp$ collisions. Adapted from Ref.~\cite{Ruiz:2015zca}.
	}
	\label{fig:}
\end{figure}

Using the analyses of Refs.~\cite{FileviezPerez:2008jbu,delAguila:2008cj,Fuks:2019clu}, and as a function of triplet scalar masses, Fig.~\ref{fig:typeIInlo_Lumi_vs_mD_LHCMulti} shows the  luminosity required to reach $5\sigma~(3\sigma)$ discovery (sensitivity) of triplet scalars produced in pairs at the LHC and subsequently decay to multi-lepton final states. (Also shown is the analogous sensitivity at a hypothetical 100 TeV $pp$ collider.) At the LHC, pairs of triplet scalars with masses in the TeV range can be discovered with the full HL-LHC data set.

%%%%%%%%%%%%%%%%%%%%%%%%%%%%%%%%%%%%%%%%%%%%%%%%%%%%%%%%%%%%%%
{\bf The Phenomenological Type III Seesaw} extends the SM field content by $n_R\geq2$ leptonic multiplets $\Sigma$ that are triplets under SU$(2)_L$. Like the Type I Seesaw, Dirac neutrino masses are generated through Yukawa couplings involving the SM Higgs field, SM lepton doublets, and $\Sigma$. The neutral component of $\Sigma$ also carries a Majorana mass, thereby triggering a high- or low-scale Seesaw depending on model assumptions. After mass diagonalization, the mass eigenstates $N,L^\pm$ can couple to charged and neutral leptons through mixing. The decomposition of gauge states into mass states can be parameterized by~\cite{Arhrib:2009mz}
\begin{equation}
    \Sigma^\pm = Y\ L^\pm + \varepsilon_T\  \ell^\pm
    \quad\text{and}\quad 
    \Sigma^0 = Y\ N + \varepsilon_T\ \nu_m,
\end{equation}
where $\vert Y\vert \sim\mathcal{O}(1)$ and $\vert \varepsilon_T\vert \ll 1$ are mixing parameters. With these couplings, Type III leptons can mediate lepton flavor violation in various low-energy observables~\cite{Abada:2007ux,Kamenik:2009cb}.

Due to their gauge couplings, triplet leptons can be produced in pairs directly from gauge interactions and through a variety of production mechanisms at proton colliders~\cite{Arhrib:2009mz,Hessler:2014ssa,Ruiz:2015zca,Cai:2017mow}. Single production in association with a SM lepton is possible through mixing.  For recent comparisons of individual channels at the LHC and beyond, see Refs~\cite{Cai:2017mow}. Typically, decays of $N$ and $L^\pm$ occurs at tree-level to SM leptons and an EW boson via mixing. However, even in the degenerate limit, $L^\pm \to N \pi^\pm$ decays are possible at one loop~\cite{Ibe:2006de,Strumia:2006db}. Such decays are allowed even in the limit of vanishing mixing. Using the analyses of Refs.~\cite{Arhrib:2009mz,Li:2009mw,delAguila:2008cj,Ruiz:2015zca}, Fig.~\ref{fig:type3_HLvsHL_LHC_Disc_vs_Mass} shows the  luminosity required to reach $5\sigma~(3\sigma)$ discovery (sensitivity) of Type III Seesaw leptons produced in pairs produced in $\sqrt{s}=14$ and 27 TeV $pp$ collisions, and which subsequently decay to multi-lepton final states.

%%%%%%%%%%%%%%%%%%%%%%%%%%%%%%%%%%%%%%%%%%%%%%%%%%%%%%%%%%%%%%

%%%%%%%%%%%%%%%%%%%%%%%%%%%%%%%%%%%%%%%%%%%%%%%%%%%%%%%%%%%%%%
{\bf The Weinberg operator at the LHC}.
Neutrinoless $\beta\beta$ decay $(0\nu\beta\beta)$ experiments are powerful probes of lepton number violation and particularly the Weinberg operator
\begin{equation}
    O^{\alpha\beta} = \left[\Phi\cdot \overline{L}^c_{\alpha} \right]\left[L_\beta \cdot \Phi\right].
\end{equation}
Here, $\Phi$ and $L$ are the usual SM Higgs and lepton doublets, with flavor indicies $\alpha,\beta$. Despite their incredible sensitivity, a limitation of $0\nu\beta\beta$ decay experiments is their flavor dependence. These experiments are only sensitive to phenomena in the $\ell_\alpha \ell_\beta = ee$ channel, and the production of same-sign lepton pairs involving muons or taus is kinematically forbidden.

As noted in Ref.~\cite{Fuks:2020zbm}, the high-$p_T$ analogue of the $0\nu\beta\beta$ process is the same-sign $W^\pm W^\pm \to \ell^\pm_\alpha \ell^\pm_\beta$ scattering channel, and is potentially accessible at the LHC. A full signal-versus-background analysis shows that the $\sqrt{s}=13$ TeV LHC with $\mathcal{L}=300$ ab$^{-1}$ (3 ab$^{-1}$) can exclude EFT cutoff scales below $\Lambda\lesssim 8.3~(11)$ TeV for a Wilson coefficient of  $C^{\alpha\beta}=1$ in the flavor channel $\ell_\alpha\ell_\beta=\mu\mu$~\cite{Fuks:2020zbm}. Anticipated sensitivity to the $\ell_\alpha\ell_\beta = ee$ and $e\mu$ channels is anticipated to be comparable, with degraded sensitivity to the $\ell_\alpha\ell_\beta = \tau\tau, \tau e$ and $\tau \mu$ to the smaller $\tau$ lepton identification. It is possible to extend this an analysis to the decays of heavy and light mesons~\cite{Atre:2005eb,Fuks:2020zbm}, thereby bringing further complementarity between high- and low-energy probes of BSM in the neutrino sector.

%%%%%%%%%%%%%%%%%%%%%%%%%%%%%%%%%%%%%%%%%%%%%%%%%%%%%%%%%%%%
\section{$Z^\prime$ model}\label{sec:zpr}

In this section, we discuss a scenario where the new physics scale is not high compared to the EW scale and therefore cannot be integrated out in low-energy tests. We focus on a $Z^\prime$ model associated with a new, anomaly-free $U(1)^\prime$ symmetry. For concreteness, we consider masses in the range $M_{Z'}=5$ MeV to few TeV and its coupling $g'<1$. 
The Lagrangian can be written as
\beq
\lag = \lag_{\rm{SM}} - \frac{1}{4}Z^{\prime \mu \nu} Z^{\prime}_{\mu\nu} + \frac{1}{2}M_{Z^\prime}^2 Z^{\prime \mu} Z^{\prime}_{\mu} + Z^{\prime}_{\mu} J_X^{\mu},
\label{eq:lagZ}
\eeq
where $J_X^\mu$ is the fermion current that couples to the $Z'$ boson.
For the sake of illustration, we take the following three cases for our benchmark studies~\cite{Heeck:2018nzc}:
\beq
(A)~Z_{B-3L_\mu},\quad(B)~Z_{B-\frac{3}{2}(L_\mu + L_\tau)},\quad(C)~Z_{B-3L_\tau}\nonumber.
\eeq

\begin{figure}[t!]
	\centering
	\subfigure[]{\includegraphics[width=0.49\textwidth]{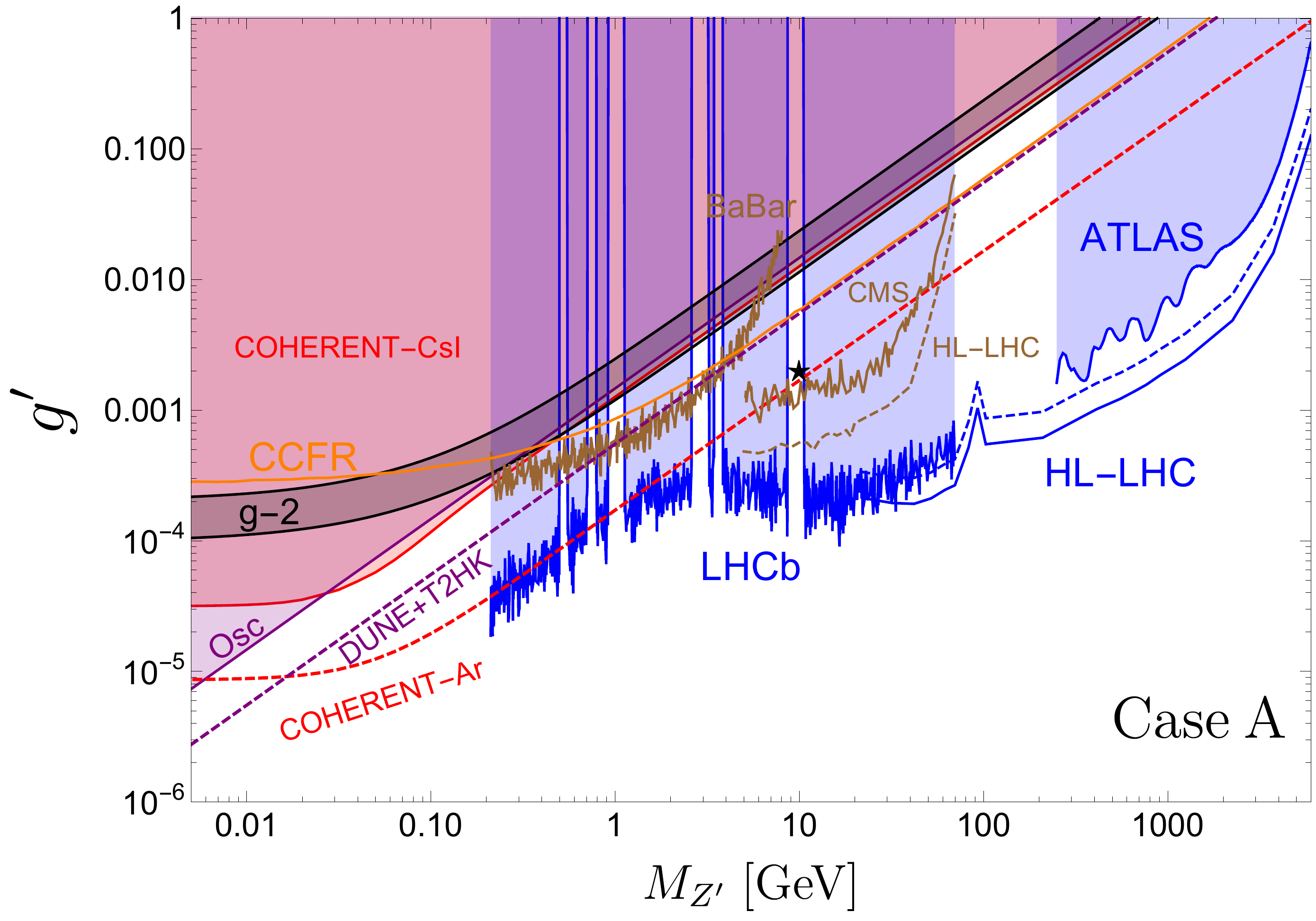}\label{fig:caseA}}
	\subfigure[]{\includegraphics[width=0.49\textwidth]{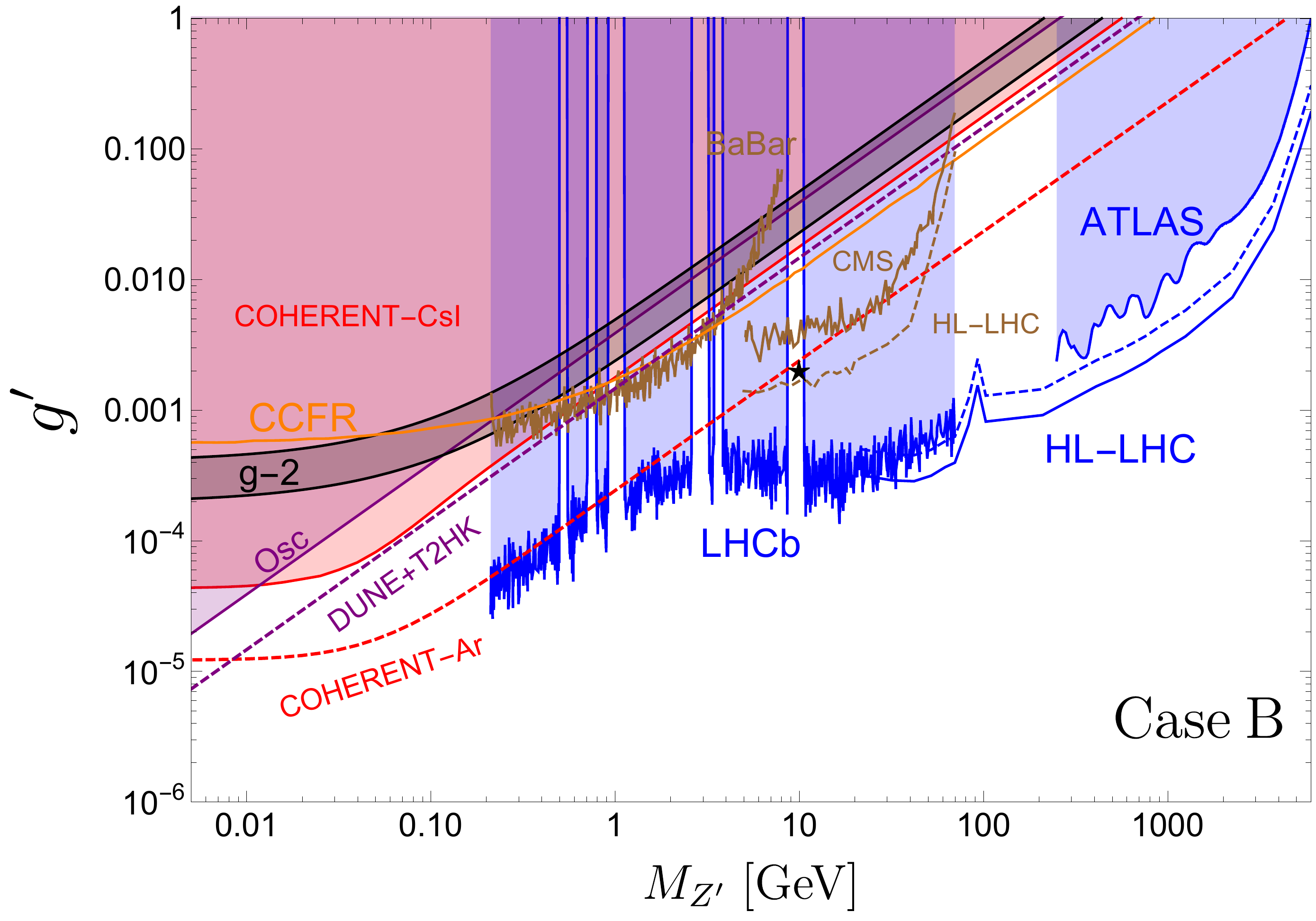}\label{fig:caseB}}
	\\
	\subfigure[]{\includegraphics[width=0.49\textwidth]{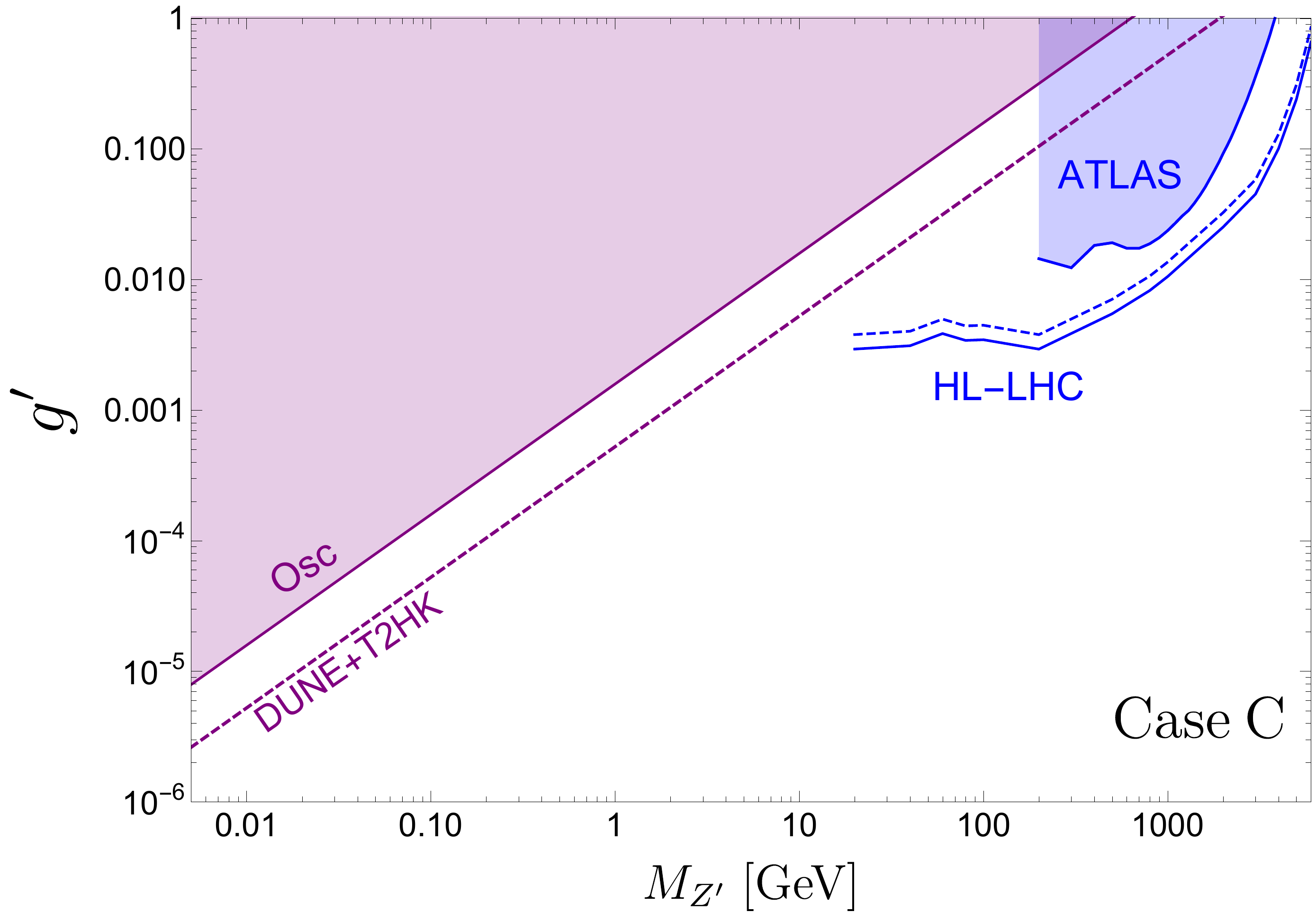}\label{fig:caseC}}
	\caption{Bounds on $g^{\prime}$ as a function of $M_{Z^\prime}$ for Cases A (upper left panel), B (upper right panel) and C~(lower panel). For details of individual experiment, see Sec.~\ref{sec:zpr}.}
	\label{fig:bounds}
\end{figure}

The parameters in these models can be constrained by both low-energy probes, like neutrino oscillation and coherent elastic neutrino-nucleus scattering, and high-energy probes like the collider experiments. The results are shown in Fig.~\ref{fig:bounds}. The red shaded areas and dashed lines correspond to the $2\sigma$ exclusion regions by using the energy spectrum from the COHERENT CsI detector~\cite{Akimov:2017ade} and the expected 2$\sigma$ limit from COHERENT with a 750~kg LAr detector and a 4-year exposure using both energy and time information~\cite{Akimov:2018ghi}, respectively. The purple shaded areas and dashed lines correspond to the $2\sigma$ bounds from a global fit to neutrino oscillation data~\cite{Esteban:2018ppq} and the expected $2\sigma$ exclusion limit from DUNE and T2HK combined, respectively. Regions above the brown curves are excluded by using $pp/e^+e^- \rightarrow \mu^+\mu^-Z^\prime$ searches at CMS~\cite{Sirunyan:2018nnz} and BaBar~\cite{TheBABAR:2016rlg} at $2\sigma$ and 90\% CL, respectively. The brown dashed curves are the $2\sigma$ expected sensitivities from the $\mu^+\mu^-Z^\prime$ channel at HL-LHC, with an integrated luminosity of 3~$\rm{ab}^{-1}$. The blue solid (dashed) curves correspond to the expected 2$\sigma$ (5$\sigma$) limit using di-muon searches for Cases A and B, and di-tau searches for Case C. The blue shaded regions in the upper panels are excluded at 90\% CL by the LHCb dark photon searches~\cite{Aaij:2019bvg} and at $2\sigma$ by the ATLAS di-muon searches~\cite{Aad:2019fac} with 139 $\rm{fb}^{-1}$. The blue area in the lower panel is excluded at $2\sigma$ by the ATLAS di-tau searches~\cite{Aaboud:2017sjh} with 36.1 $\rm{fb}^{-1}$. The orange curves correspond to the 2$\sigma$ limit from CCFR~\cite{Mishra:1991bv,Altmannshofer:2014pba}, and the black bands show the 2$\sigma$ allowed regions that explain the discrepancy in the anomalous magnetic moment of the muon ($\Delta a_\mu = (29\pm 9 )\times 10^{-10}$~\cite{Jegerlehner:2009ry}).

The low- and high-energy probes are complementary to search for a wide range of $Z^\prime$ masses. For $Z'$ masses above 10~GeV, the ATLAS and CMS experiments at HL-LHC have the best sensitivity regardless of the flavor structure of the model. For masses between 0.01~GeV$-10$~GeV, current LHCb  data and future COHERENT data have the best sensitivity unless the $Z'$ couplings to the first and second generation leptons are suppressed, in which case neutrino oscillation experiments have the best sensitivity. For $Z'$ masses between about 5~MeV$-20$~MeV, DUNE and T2HK have the best sensitivity. 

%%%%%%%%%%%%%%%%%%%%%%%%%%%%%%%%%%%%%%%%%%%%%%%%%%%%%%%%%%%%%
\section{$\nu$SMEFT / SMNEFT}

\begin{figure}[t!]
	\centering
\subfigure[]{\includegraphics[width=0.45\textwidth]{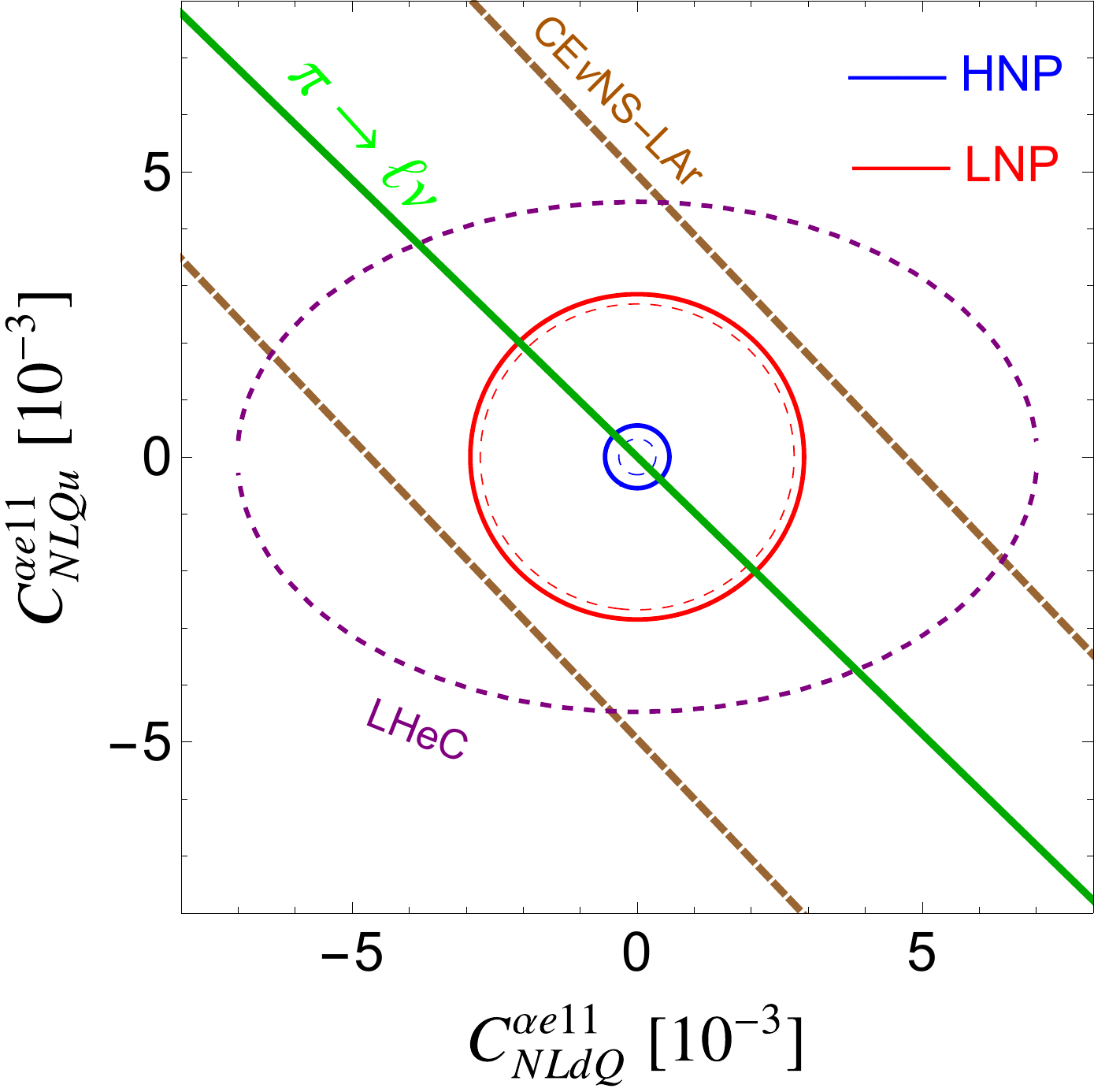}}
\subfigure[]{\includegraphics[width=0.45\textwidth]{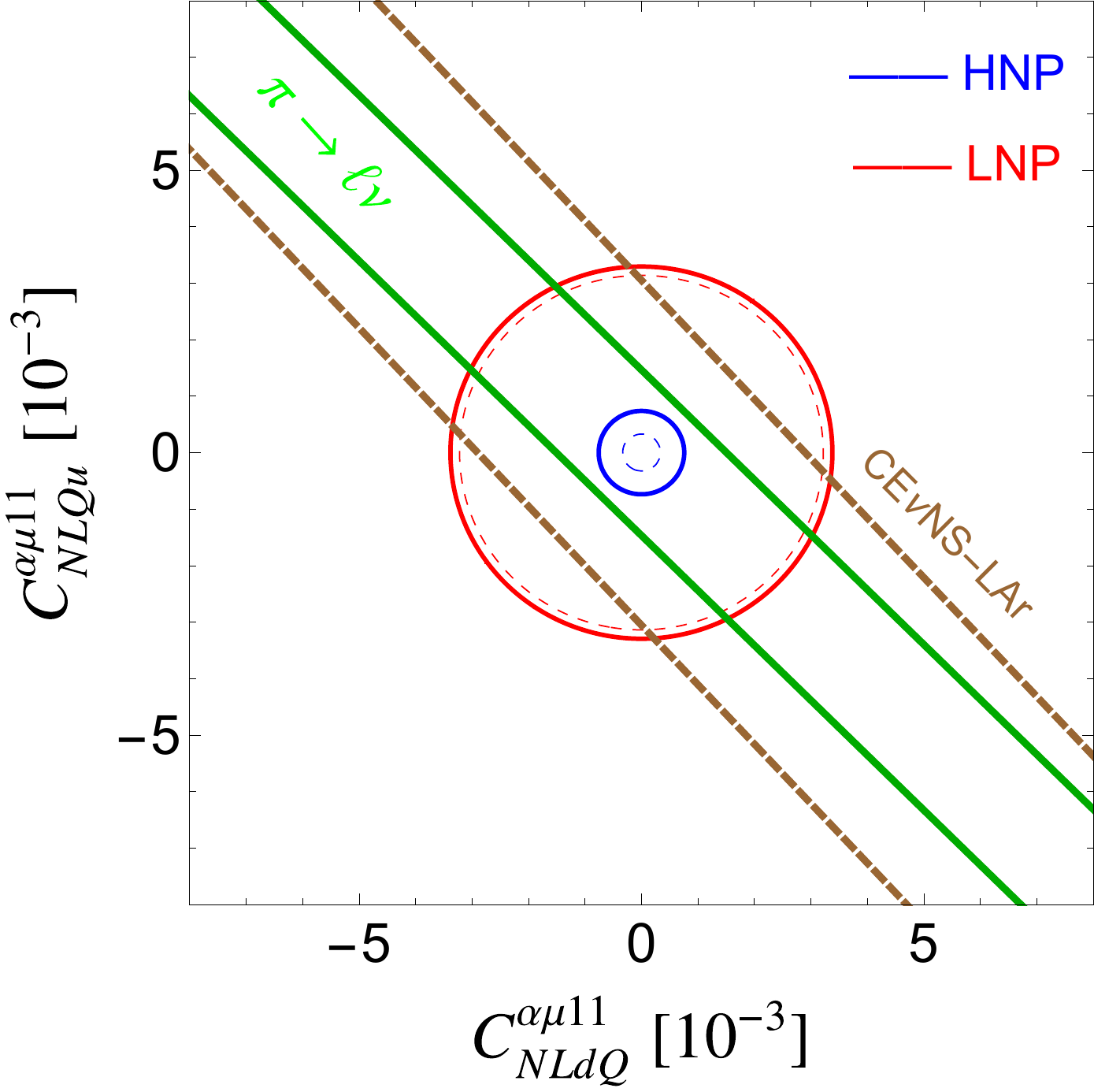}}
\\
\subfigure[]{\includegraphics[width=0.45\textwidth]{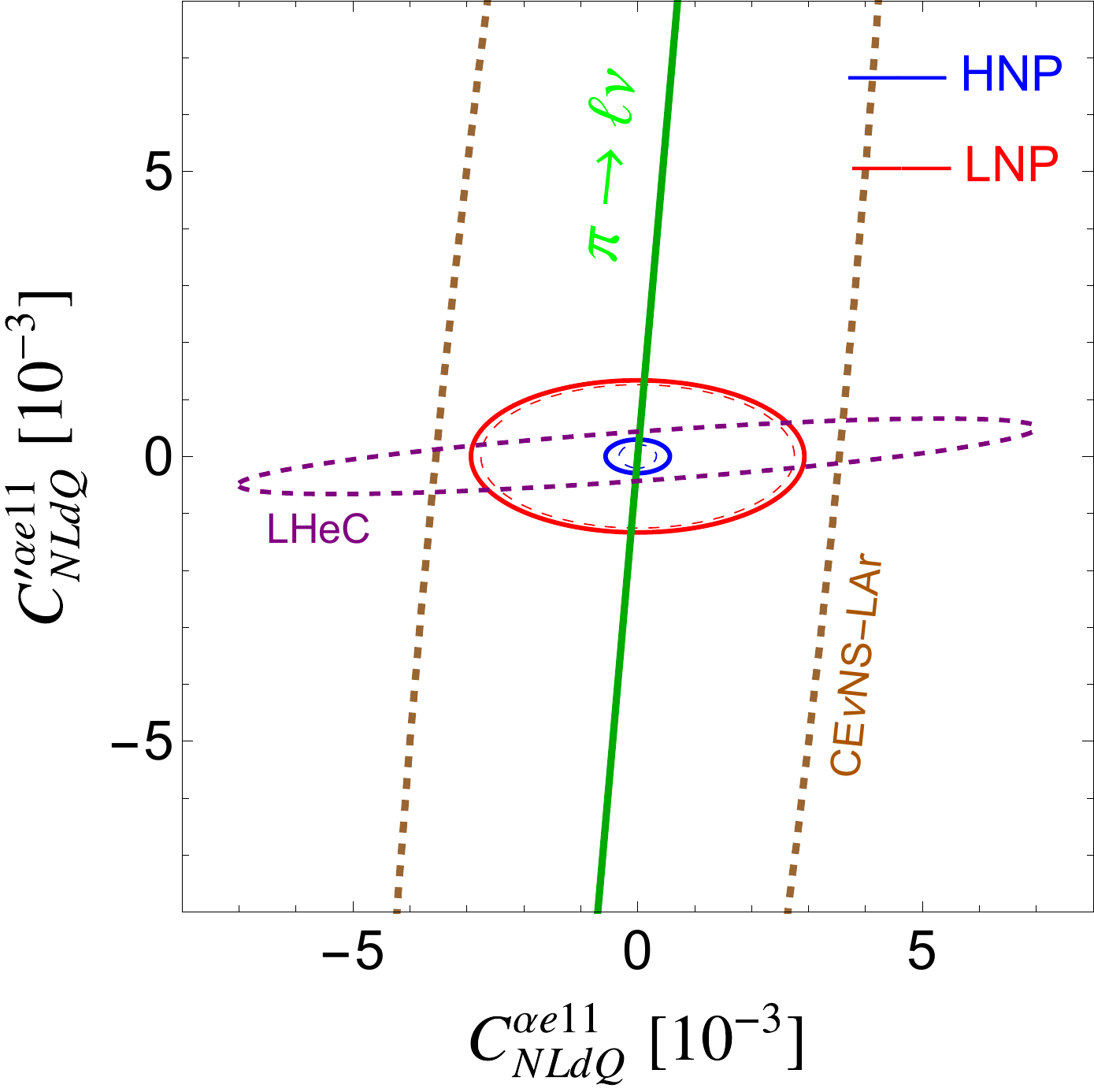}}
\subfigure[]{\includegraphics[width=0.45\textwidth]{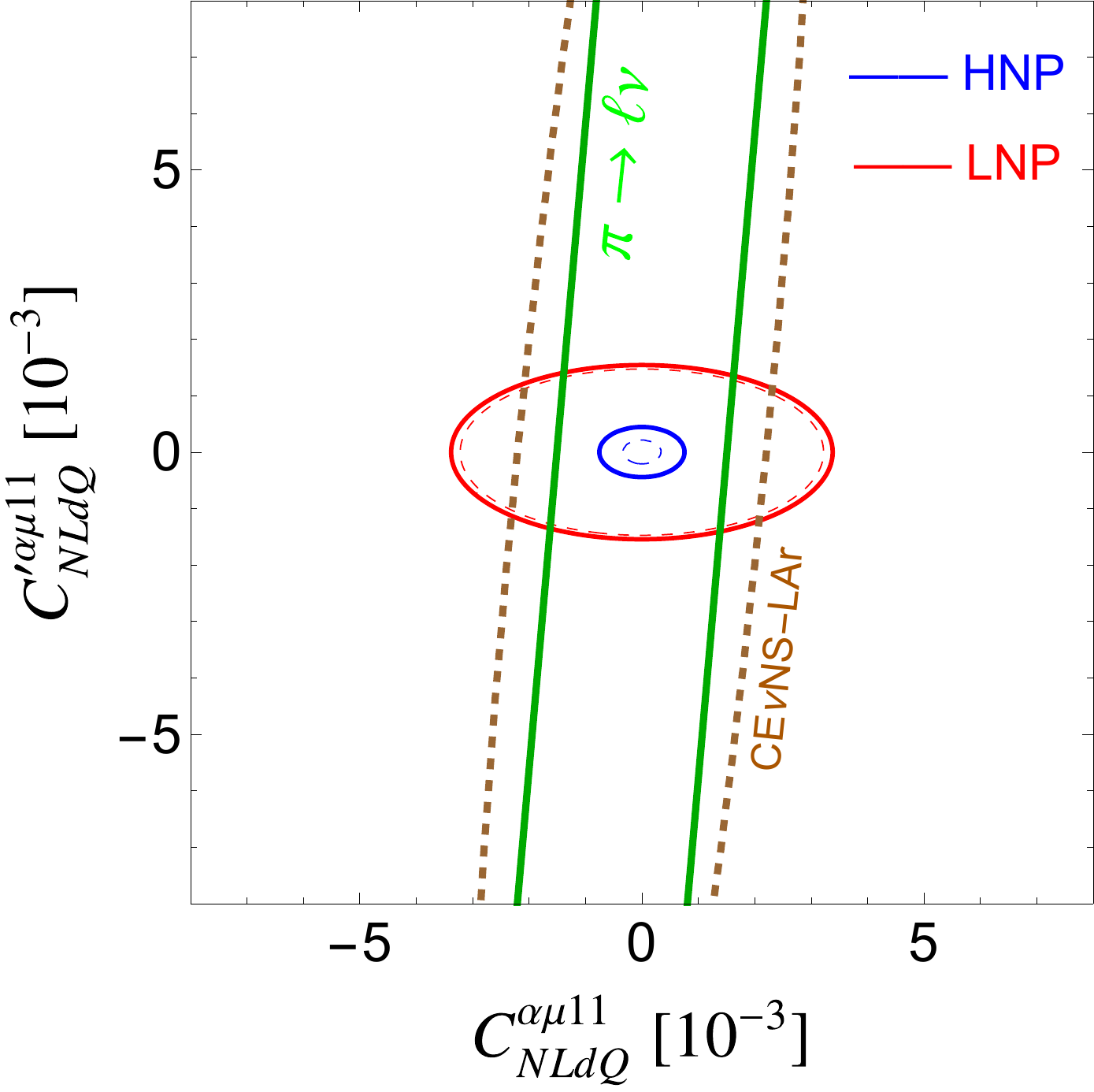}}
	\caption{The 90\%~C.L. allowed regions in the $C_{NLdQ}$-$C_{NLQu}$ planes (upper panels) and $C_{NLdQ}$-$C^\prime_{NLdQ}$ planes (lower panels) at 1~TeV with $\ell=e$ flavor (left) and $\ell=\mu$ flavor (right). The green lines (overlapping in the left panels) correspond to the bounds from pion decay with the third parameter set to zero to break the degeneracy. The red (blue) solid contours show current  LHC  searches with $L = 139\,\text{fb}^{-1}$ for the low-scale new physics LNP (high-scale new physics HNP) case. The brown dashed lines correspond to the projected bounds from the future LAr COHERENT experiment, with $C_{NLQu}$ is set to zero in the lower panels to obtain meaningful bounds. The red and blue dashed contours are the projected bounds from HL-LHC with 3~$\text{ab}^{-1}$ of data for the LNP and HNP case, respectively. The dashed purple contours in the left panels correspond to the projected bounds from LHeC with 3~$\text{ab}^{-1}$.}
	\label{fig:Coll_Scalar_EFT}
\end{figure} 

When the new physics scale is much higher than the weak scale, an EFT framework that is consistent with the SM gauge symmetries can be used above the weak scale. To parameterize scenarios where the mass of a heavy neutrino state is comparable to the EW scale but the scale of GNI is much larger than both, it is possible to extend the SM by right-handed neutrinos and build an effective field theory based on the given field content.  
The framework, known as $\nu$SMEFT or SMNEFT, reduces to the familiar SMEFT when heavy neutrinos are integrated out of the theory~\cite{delAguila:2008ir,Bhattacharya:2015vja,Liao:2016qyd}. Thus, $\nu$SMEFT is a broader framework than SMEFT and can generate GNI at low energies. We focus on four operators: 
\begin{enumerate}
	\item $O^{\alpha\beta\gamma\delta}_{NLQu} =  (\overline{N}_{\alpha} L_{\beta}^j)(\overline{Q}^j_\gamma u_\delta)$\,,
	\item $O^{\alpha\beta\gamma\delta}_{NLdQ} = (\overline{N}_{\alpha} L^j_{\beta})\epsilon_{jk}(\overline{d}_\gamma Q^k_\delta)$\,,
	\item $O^{\prime\alpha\beta\gamma\delta}_{NLdQ} = (\overline{N}_{\alpha}\sigma_{\mu\nu} L^j_{\beta})\epsilon_{jk}(\overline{d}_\gamma \sigma^{\mu\nu} Q^k_\delta)$\,,
	\item $O^{}_{HNe} = \left(i \tilde{\varphi}^\dagger D_\mu\varphi\right)\overline{N}\sigma^\mu e_R$\,,
\end{enumerate}
where the fields are written in two-component spinors. $L$ and $Q$ are the left-handed lepton and quark doublet, respectively, 
$\varphi$ is the left-handed Higgs doublet,
and $N$ is the right-handed neutrino state. Here, $\sigma^{\mu\nu} = \frac{i}{2}[\sigma^\mu\overline{\sigma}^\nu - \sigma^\nu\overline{\sigma}^\mu]$, with $\sigma^\mu=(\mathbb{1}, \vec{\mathbf{\sigma}})$ and $\bar{\sigma}^\mu=(\mathbb{1}, -\vec{\mathbf{\sigma}})$. RG running is needed to combine the results from various energy scales and it has been calculated in Refs.~\cite{Datta:2020ocb, Chala:2020pbn, Han:2020pff, Datta:2021akg}. 
Generally there are degeneracies between $\nu$SMEFT Wilson coefficients at low-energy observables due to the running effects. High-energy measurements can break the degeneracies. Low-energy probes and high-energy colliders are complementary to probe various type of interactions as shown in Fig.~\ref{fig:Coll_Scalar_EFT}.

\begin{figure}[t!]
\centering
\subfigure[]{\includegraphics[width=.48\textwidth]{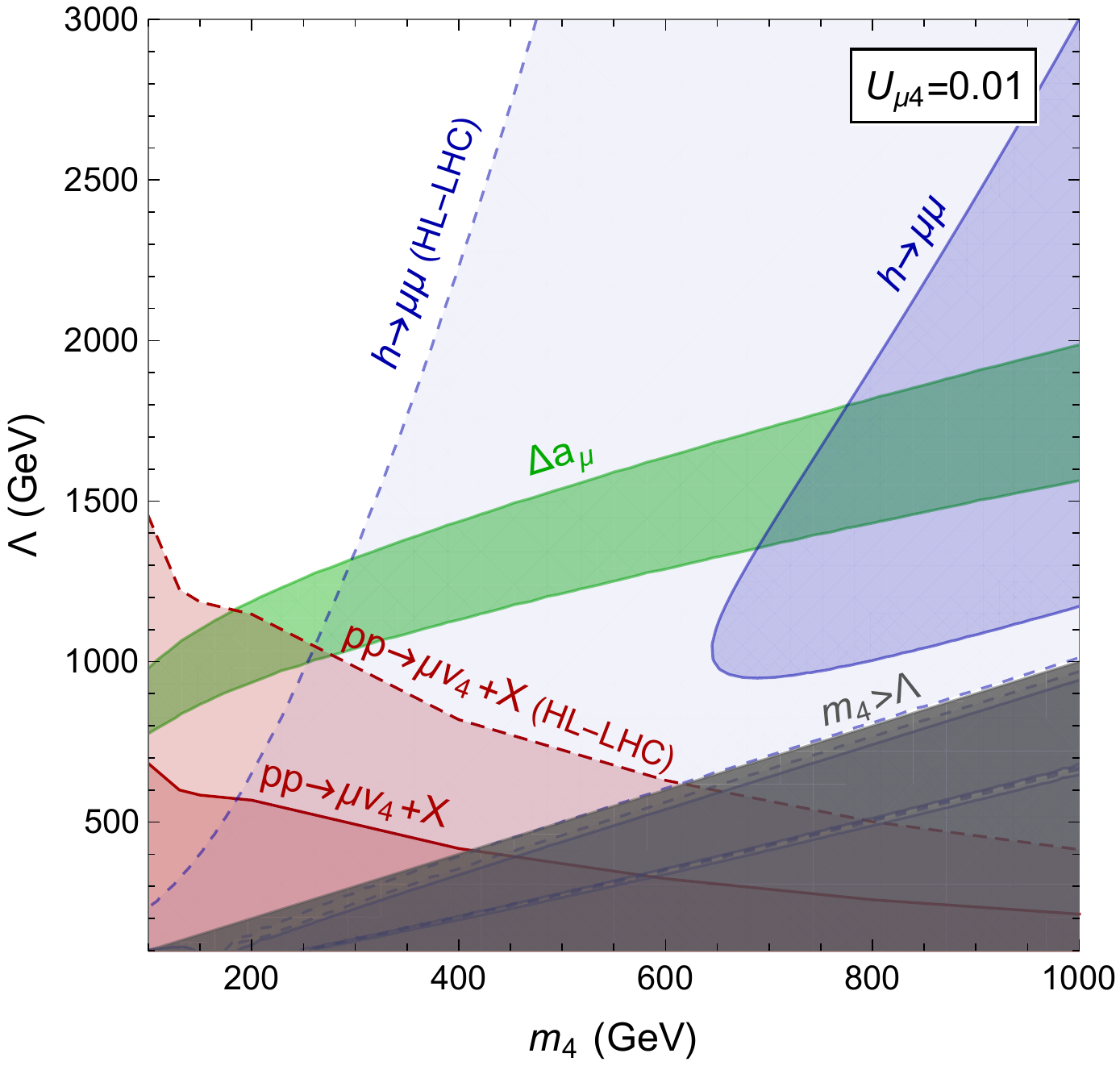} \label{fig:vSMEFTg2_CHNePlotv2}}
\subfigure[]{\includegraphics[width=.48\textwidth]{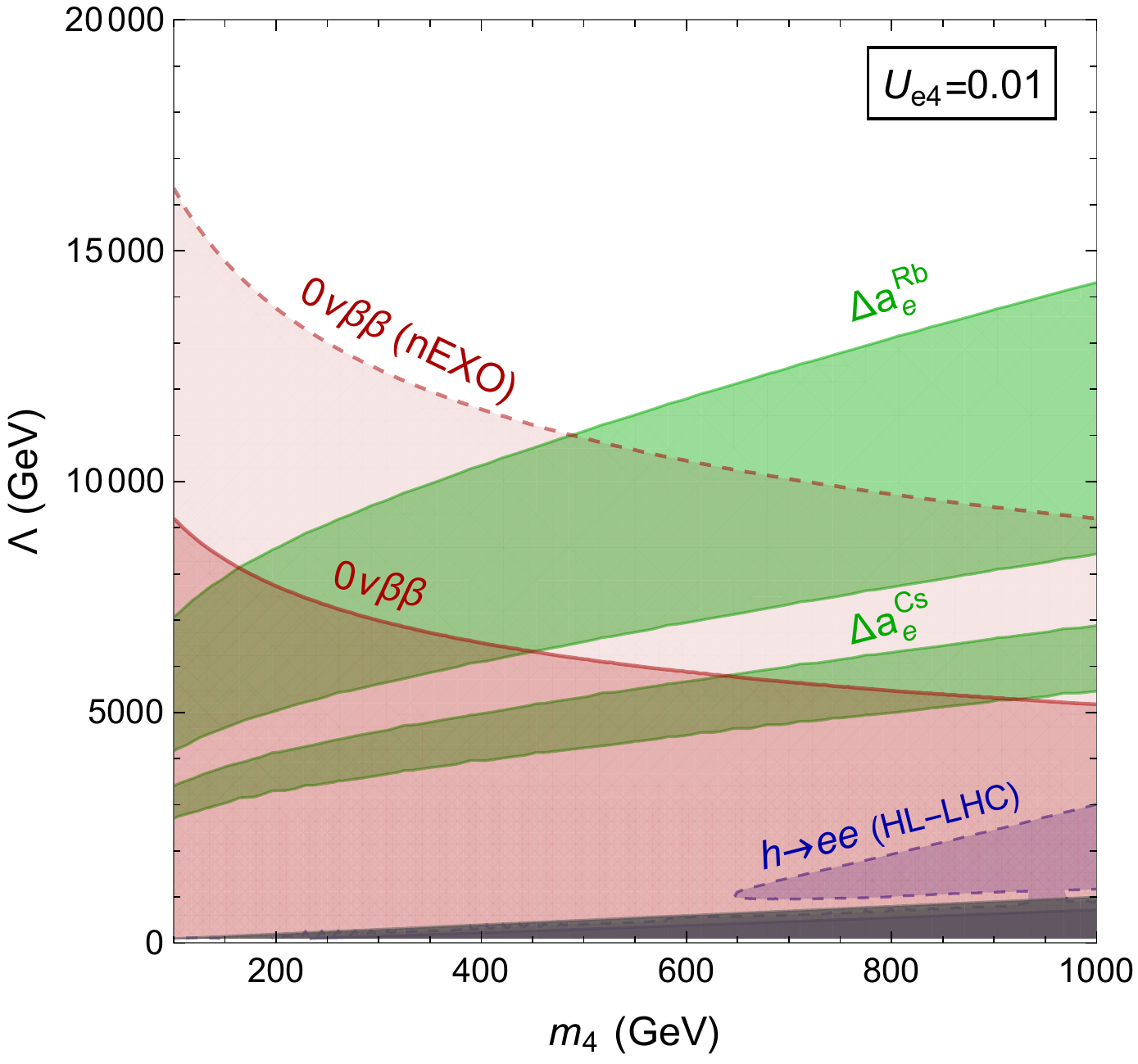} \label{fig:vSMEFTg2_electron_plotv2}}
	\caption{
(a) For a Wilson coefficient of unity and active-sterile mixing $\vert U_{\mu 4}\vert^2 = 10^{-4}$, the region in $(\Lambda,m_4)$ space that can account for the measured value of $\Delta a_{\mu}$ (green band); current exclusions due to direct searches for sterile neutrinos (dark red) and $H\to\mu^+\mu^-$ decays (dark blue) at $\sqrt{s}=13$ TeV; the anticipated sensitivity at the HL-LHC in the same channels (light red and light blue); and exclusions due to requiring $m_4 < \Lambda$.
(b) Same as (a) but for $\ell=e$ with present constraints (dark red) as well as anticipated sensitivity  from $0\nu\beta\beta$ experiments (light red) and searches for $H\to e^+e^-$ at the HL-LHC. Figure adapted from Ref.~\cite{Cirigliano:2021peb}.
}
	\label{fig:vsmeft_gm2}
\end{figure} 

If EW-scale sterile neutrinos contribute to the anomalous magnetic moments of charged leptons, then it is possible to parameterize their impact with the $\nu$SMEFT framework~\cite{Cirigliano:2021peb}. Notably, only one operator at dimension $d=6$ can account for the difference between recent measurements and predictions for the muon's anomalous magnetic moment, $\Delta a_{\mu}$. In particular,  with the exception of operator $O^{}_{HNe}$ above, all other operators at $d=6$ either generate too small or too large $\Delta a_{\mu}$~\cite{Cirigliano:2021peb}.
Moreover, the measured value of $\Delta a_{\mu}$ is sufficiently large that the allowed band of $(C_{HNe}/\Lambda)$ and $m_N$, with realistic active-sterile mixing is accessible at the LHC. (Here, $C_{HNe}$ is the Wilson coefficient for $O^{}_{HNe}$, $\Lambda$ is the EFT cutoff scale, $m_4$ is the mass of the heavy mass eigenstate $N$.)

For a Wilson coefficient of unity and active-sterile mixing $\vert U_{\mu 4}\vert^2 = 10^{-4}$,
Fig.~\ref{fig:vSMEFTg2_CHNePlotv2} shows the region in $(\Lambda,m_4)$ space that can account for the measured value of $\Delta a_{\mu}$ (green band); 
current exclusions due to direct searches for sterile neutrinos with $\mathcal{L}\approx 36$ fb$^{-1}$ (dark red)~\cite{} and $H\to\mu^+\mu^-$ decays with $\mathcal{L}\approx 137$ fb$^{-1}$ (dark blue)~\cite{CMS:2020xwi,ATLAS:2020fzp}
at $\sqrt{s}=13$ TeV; the anticipated sensitivity at the HL-LHC in the same channels (light red and light blue)~\cite{Cepeda:2019klc}; and exclusions due to requiring $m_4 < \Lambda$. By the end of the HL-LHC program, nearly the entire allowed parameter space can be explored at the HL-LHC~\cite{Cirigliano:2021peb}.

A disagreement between experimental determinations of the electron's anomalous magnetic moment using Cesium~\cite{Parker:2018vye} and Rubidium~\cite{Morel:2020dww} also exists. While inaccessible at the HL-LHC, the nEXO neutrinoless $\beta\beta$ $(0\nu\beta\beta)$ decay experiment~\cite{EXO-200:2019rkq} will be able to explore much of the allowed $(\Lambda,m_4)$ space.
This is shown in Fig.~\ref{fig:vSMEFTg2_electron_plotv2}, 
where the allowed space for $\Delta a_e$ is shown as are 
present constraints (dark red) as well as anticipated sensitivity  from $0\nu\beta\beta$ experiments (light red) and searches for $H\to e^+e^-$ at the HL-LHC based on current exclusions~\cite{ATLAS:2019old}. Figure adapted from Ref.~\cite{Cirigliano:2021peb}.

Finally, other studies in the $\nu$SMEFT framework report sensitivity at the LHC to individual operators at the LHC~\cite{delAguila:2008ir,Aparici:2009fh,Bhattacharya:2015vja,Duarte:2015iba,Ruiz:2017nip,Cottin:2021lzz,Beltran:2021hpq} as well as proposed lepton colliders~\cite{Duarte:2016miz,Duarte:2016caz,Duarte:2018kiv,Zapata:2022qwo}.

%%%%%%%%%%%%%%%%%%%%%%%%%%%%%%%%%%%%%%%%%%%%%%%%%%%%%%%%%%%%%%%%%%5
%%%%%%%%%%%%%%%%%%%%%%%%%%%%%%%%%%%%%%%%%%%%%%%%%%%%%%%%%%%%%%%%%%5

\section{Executive summary}
Through a few phenomenological examples, we reiterate that there is significant complementarity between low-energy experiments (such as DUNE) and high-energy collider experiments (such as the LHC) in exploring new physics associated with neutrino properties and their mass-generation mechanisms. Signals of new  physics in the two energy regimes may be correlated with each other in the same  underlying theory. We demonstrate the complementary nature by presenting the physics reaches for the Seesaw models of Type I, II and III, and for general neutrino interactions in an effective field theory framework, and in a $Z'$ model. It is crucially important to seek complementary signals to establish a consistent picture of the underlying physics associated with the mechanisms responsible for generating neutrino masses.

%%%%%%%%%%%%%%%%%%%%%%%%%%%%%%%%%%%%%%%%%%%%%%%%%%%%%%%%%%%%%%%%%%

\vskip 0.5cm 
\noindent
{\bf Acknowledgments}\\
The work of T.H. was supported in part by the U.S.~Department of Energy (DoE) under grant 
No.~DE-SC0007914 
and in part by the PITT PACC. J.L. is supported by the National Natural Science Foundation of China under Grant No. 11905299 and Guangdong Basic and Applied Basic Research Foundation under Grant No. 2020A1515011479. H.L. is supported by ISF, BSF and Azrieli foundation.
 D.M. is supported by the DoE under Grant No. DE-SC-0010504.  D. M. thanks KITP, Santa Barbara for its hospitality and support via the NSF under Grant No. PHY-1748958 during the completion of this work. R.R. acknowledges the support of the Polska Akademia Nauk (grant agreement PAN.BFD.S.BDN. 613. 022. 2021 - PASIFIC 1, POPSICLE). This work has received funding from the European Union's Horizon 2020 research and innovation program under the Sk{\l}odowska-Curie grant agreement No.  847639 and from the Polish Ministry of Education and Science.

%%%%%%%%%%%%%%%%%%%%%%%%%%%%%%%%%%%%%%%%%%%%%%%%%%%%%%%%%%%%%%%%%%

\bibliographystyle{JHEPMod}
\bibliography{nuComplementarity}

\providecommand{\href}[2]{#2}\begingroup\raggedright\begin{thebibliography}{10}

\bibitem{Ma:1998dn}
E.~Ma, \textit{{Pathways to naturally small neutrino masses}},
  \href{https://doi.org/10.1103/PhysRevLett.81.1171}{\textit{Phys. Rev. Lett.}
  {\bfseries 81} (1998) 1171--1174},
  [\href{https://arxiv.org/abs/hep-ph/9805219}{{\ttfamily hep-ph/9805219}}].

\bibitem{Deppisch:2015qwa}
F.~F. Deppisch, P.~S. Bhupal~Dev and A.~Pilaftsis, \textit{{Neutrinos and
  Collider Physics}},
  \href{https://doi.org/10.1088/1367-2630/17/7/075019}{\textit{New J. Phys.}
  {\bfseries 17} (2015) 075019},
  [\href{https://arxiv.org/abs/1502.06541}{{\ttfamily 1502.06541}}].

\bibitem{Cai:2017jrq}
Y.~Cai, J.~Herrero-Garc\'\i{}a, M.~A. Schmidt, A.~Vicente and R.~R. Volkas,
  \textit{{From the trees to the forest: a review of radiative neutrino mass
  models}}, \href{https://doi.org/10.3389/fphy.2017.00063}{\textit{Front. in
  Phys.} {\bfseries 5} (2017) 63},
  [\href{https://arxiv.org/abs/1706.08524}{{\ttfamily 1706.08524}}].

\bibitem{Cai:2017mow}
Y.~Cai, T.~Han, T.~Li and R.~Ruiz, \textit{{Lepton Number Violation: Seesaw
  Models and Their Collider Tests}},
  \href{https://doi.org/10.3389/fphy.2018.00040}{\textit{Front. in Phys.}
  {\bfseries 6} (2018) 40}, [\href{https://arxiv.org/abs/1711.02180}{{\ttfamily
  1711.02180}}].

\bibitem{Han:2019zkz}
T.~Han, J.~Liao, H.~Liu and D.~Marfatia, \textit{{Nonstandard neutrino
  interactions at COHERENT, DUNE, T2HK and LHC}},
  \href{https://doi.org/10.1007/JHEP11(2019)028}{\textit{JHEP} {\bfseries 11}
  (2019) 028}, [\href{https://arxiv.org/abs/1910.03272}{{\ttfamily
  1910.03272}}].

\bibitem{Han:2020pff}
T.~Han, J.~Liao, H.~Liu and D.~Marfatia, \textit{{Scalar and tensor neutrino
  interactions}}, \href{https://doi.org/10.1007/JHEP07(2020)207}{\textit{JHEP}
  {\bfseries 07} (2020) 207},
  [\href{https://arxiv.org/abs/2004.13869}{{\ttfamily 2004.13869}}].

\bibitem{Cirigliano:2021peb}
V.~Cirigliano, W.~Dekens, J.~de~Vries, K.~Fuyuto, E.~Mereghetti and R.~Ruiz,
  \textit{{Leptonic anomalous magnetic moments in \ensuremath{\nu} SMEFT}},
  \href{https://doi.org/10.1007/JHEP08(2021)103}{\textit{JHEP} {\bfseries 08}
  (2021) 103}, [\href{https://arxiv.org/abs/2105.11462}{{\ttfamily
  2105.11462}}].

\bibitem{delAguila:2008cj}
F.~del Aguila and J.~A. Aguilar-Saavedra, \textit{{Distinguishing seesaw models
  at LHC with multi-lepton signals}},
  \href{https://doi.org/10.1016/j.nuclphysb.2008.12.029}{\textit{Nucl. Phys. B}
  {\bfseries 813} (2009) 22--90},
  [\href{https://arxiv.org/abs/0808.2468}{{\ttfamily 0808.2468}}].

\bibitem{Atre:2009rg}
A.~Atre, T.~Han, S.~Pascoli and B.~Zhang, \textit{{The Search for Heavy
  Majorana Neutrinos}},
  \href{https://doi.org/10.1088/1126-6708/2009/05/030}{\textit{JHEP} {\bfseries
  05} (2009) 030}, [\href{https://arxiv.org/abs/0901.3589}{{\ttfamily
  0901.3589}}].

\bibitem{Keung:1983uu}
W.-Y. Keung and G.~Senjanovic, \textit{{Majorana Neutrinos and the Production
  of the Right-handed Charged Gauge Boson}},
  \href{https://doi.org/10.1103/PhysRevLett.50.1427}{\textit{Phys. Rev. Lett.}
  {\bfseries 50} (1983) 1427}.

\bibitem{Datta:1993nm}
A.~Datta, M.~Guchait and A.~Pilaftsis, \textit{{Probing lepton number violation
  via majorana neutrinos at hadron supercolliders}},
  \href{https://doi.org/10.1103/PhysRevD.50.3195}{\textit{Phys. Rev. D}
  {\bfseries 50} (1994) 3195--3203},
  [\href{https://arxiv.org/abs/hep-ph/9311257}{{\ttfamily hep-ph/9311257}}].

\bibitem{Dev:2013wba}
P.~S.~B. Dev, A.~Pilaftsis and U.-k. Yang, \textit{{New Production Mechanism
  for Heavy Neutrinos at the LHC}},
  \href{https://doi.org/10.1103/PhysRevLett.112.081801}{\textit{Phys. Rev.
  Lett.} {\bfseries 112} (2014) 081801},
  [\href{https://arxiv.org/abs/1308.2209}{{\ttfamily 1308.2209}}].

\bibitem{Alva:2014gxa}
D.~Alva, T.~Han and R.~Ruiz, \textit{{Heavy Majorana neutrinos from $W\gamma$
  fusion at hadron colliders}},
  \href{https://doi.org/10.1007/JHEP02(2015)072}{\textit{JHEP} {\bfseries 02}
  (2015) 072}, [\href{https://arxiv.org/abs/1411.7305}{{\ttfamily 1411.7305}}].

\bibitem{Degrande:2016aje}
C.~Degrande, O.~Mattelaer, R.~Ruiz and J.~Turner, \textit{{Fully-Automated
  Precision Predictions for Heavy Neutrino Production Mechanisms at Hadron
  Colliders}}, \href{https://doi.org/10.1103/PhysRevD.94.053002}{\textit{Phys.
  Rev. D} {\bfseries 94} (2016) 053002},
  [\href{https://arxiv.org/abs/1602.06957}{{\ttfamily 1602.06957}}].

\bibitem{Dicus:1991wj}
D.~A. Dicus and P.~Roy, \textit{{Supercollider signatures and correlations of
  heavy neutrinos}},
  \href{https://doi.org/10.1103/PhysRevD.44.1593}{\textit{Phys. Rev. D}
  {\bfseries 44} (1991) 1593--1596}.

\bibitem{Hessler:2014ssa}
A.~G. Hessler, A.~Ibarra, E.~Molinaro and S.~Vogl, \textit{{Impact of the Higgs
  boson on the production of exotic particles at the LHC}},
  \href{https://doi.org/10.1103/PhysRevD.91.115004}{\textit{Phys. Rev. D}
  {\bfseries 91} (2015) 115004},
  [\href{https://arxiv.org/abs/1408.0983}{{\ttfamily 1408.0983}}].

\bibitem{Ruiz:2017yyf}
R.~Ruiz, M.~Spannowsky and P.~Waite, \textit{{Heavy neutrinos from gluon
  fusion}}, \href{https://doi.org/10.1103/PhysRevD.96.055042}{\textit{Phys.
  Rev. D} {\bfseries 96} (2017) 055042},
  [\href{https://arxiv.org/abs/1706.02298}{{\ttfamily 1706.02298}}].

\bibitem{Dicus:1991fk}
D.~A. Dicus, D.~D. Karatas and P.~Roy, \textit{{Lepton nonconservation at
  supercollider energies}},
  \href{https://doi.org/10.1103/PhysRevD.44.2033}{\textit{Phys. Rev. D}
  {\bfseries 44} (1991) 2033--2037}.

\bibitem{Fuks:2020att}
B.~Fuks, J.~Neundorf, K.~Peters, R.~Ruiz and M.~Saimpert, \textit{{Majorana
  neutrinos in same-sign $W^\pm W^\pm$ scattering at the LHC: Breaking the TeV
  barrier}}, \href{https://doi.org/10.1103/PhysRevD.103.055005}{\textit{Phys.
  Rev. D} {\bfseries 103} (2021) 055005},
  [\href{https://arxiv.org/abs/2011.02547}{{\ttfamily 2011.02547}}].

\bibitem{Pascoli:2018heg}
S.~Pascoli, R.~Ruiz and C.~Weiland, \textit{{Heavy neutrinos with dynamic jet
  vetoes: multilepton searches at $ \sqrt{s}=14 $ , 27, and 100 TeV}},
  \href{https://doi.org/10.1007/JHEP06(2019)049}{\textit{JHEP} {\bfseries 06}
  (2019) 049}, [\href{https://arxiv.org/abs/1812.08750}{{\ttfamily
  1812.08750}}].

\bibitem{Pascoli:2018rsg}
S.~Pascoli, R.~Ruiz and C.~Weiland, \textit{{Safe Jet Vetoes}},
  \href{https://doi.org/10.1016/j.physletb.2018.08.060}{\textit{Phys. Lett. B}
  {\bfseries 786} (2018) 106--113},
  [\href{https://arxiv.org/abs/1805.09335}{{\ttfamily 1805.09335}}].

\bibitem{Fernandez-Martinez:2016lgt}
E.~Fernandez-Martinez, J.~Hernandez-Garcia and J.~Lopez-Pavon, \textit{{Global
  constraints on heavy neutrino mixing}},
  \href{https://doi.org/10.1007/JHEP08(2016)033}{\textit{JHEP} {\bfseries 08}
  (2016) 033}, [\href{https://arxiv.org/abs/1605.08774}{{\ttfamily
  1605.08774}}].

\bibitem{Han:2012vk}
T.~Han, I.~Lewis, R.~Ruiz and Z.-g. Si, \textit{{Lepton Number Violation and
  $W^\prime$ Chiral Couplings at the LHC}},
  \href{https://doi.org/10.1103/PhysRevD.87.035011}{\textit{Phys. Rev. D}
  {\bfseries 87} (2013) 035011},
  [\href{https://arxiv.org/abs/1211.6447}{{\ttfamily 1211.6447}}].

\bibitem{Ruiz:2017nip}
R.~Ruiz, \textit{{Lepton Number Violation at Colliders from Kinematically
  Inaccessible Gauge Bosons}},
  \href{https://doi.org/10.1140/epjc/s10052-017-4950-2}{\textit{Eur. Phys. J.
  C} {\bfseries 77} (2017) 375},
  [\href{https://arxiv.org/abs/1703.04669}{{\ttfamily 1703.04669}}].

\bibitem{Ruiz:2020cjx}
R.~Ruiz, \textit{{Quantitative study on helicity inversion in Majorana neutrino
  decays at the LHC}},
  \href{https://doi.org/10.1103/PhysRevD.103.015022}{\textit{Phys. Rev. D}
  {\bfseries 103} (2021) 015022},
  [\href{https://arxiv.org/abs/2008.01092}{{\ttfamily 2008.01092}}].

\bibitem{Fuks:2019clu}
B.~Fuks, M.~Nemev\v{s}ek and R.~Ruiz, \textit{{Doubly Charged Higgs Boson
  Production at Hadron Colliders}},
  \href{https://doi.org/10.1103/PhysRevD.101.075022}{\textit{Phys. Rev. D}
  {\bfseries 101} (2020) 075022},
  [\href{https://arxiv.org/abs/1912.08975}{{\ttfamily 1912.08975}}].

\bibitem{Nemevsek:2016enw}
M.~Nemev\v{s}ek, F.~Nesti and J.~C. Vasquez, \textit{{Majorana Higgses at
  colliders}}, \href{https://doi.org/10.1007/JHEP04(2017)114}{\textit{JHEP}
  {\bfseries 04} (2017) 114},
  [\href{https://arxiv.org/abs/1612.06840}{{\ttfamily 1612.06840}}].

\bibitem{FileviezPerez:2008jbu}
P.~Fileviez~Perez, T.~Han, G.-y. Huang, T.~Li and K.~Wang, \textit{{Neutrino
  Masses and the CERN LHC: Testing Type II Seesaw}},
  \href{https://doi.org/10.1103/PhysRevD.78.015018}{\textit{Phys. Rev. D}
  {\bfseries 78} (2008) 015018},
  [\href{https://arxiv.org/abs/0805.3536}{{\ttfamily 0805.3536}}].

\bibitem{Ruiz:2015zca}
R.~Ruiz, \textit{{QCD Corrections to Pair Production of Type III Seesaw Leptons
  at Hadron Colliders}},
  \href{https://doi.org/10.1007/JHEP12(2015)165}{\textit{JHEP} {\bfseries 12}
  (2015) 165}, [\href{https://arxiv.org/abs/1509.05416}{{\ttfamily
  1509.05416}}].

\bibitem{Arhrib:2009mz}
A.~Arhrib, B.~Bajc, D.~K. Ghosh, T.~Han, G.-Y. Huang, I.~Puljak et~al.,
  \textit{{Collider Signatures for Heavy Lepton Triplet in Type I+III Seesaw}},
  \href{https://doi.org/10.1103/PhysRevD.82.053004}{\textit{Phys. Rev. D}
  {\bfseries 82} (2010) 053004},
  [\href{https://arxiv.org/abs/0904.2390}{{\ttfamily 0904.2390}}].

\bibitem{Abada:2007ux}
A.~Abada, C.~Biggio, F.~Bonnet, M.~B. Gavela and T.~Hambye, \textit{{Low energy
  effects of neutrino masses}},
  \href{https://doi.org/10.1088/1126-6708/2007/12/061}{\textit{JHEP} {\bfseries
  12} (2007) 061}, [\href{https://arxiv.org/abs/0707.4058}{{\ttfamily
  0707.4058}}].

\bibitem{Kamenik:2009cb}
J.~F. Kamenik and M.~Nemevsek, \textit{{Lepton flavor violation in type I + III
  seesaw}}, \href{https://doi.org/10.1088/1126-6708/2009/11/023}{\textit{JHEP}
  {\bfseries 11} (2009) 023},
  [\href{https://arxiv.org/abs/0908.3451}{{\ttfamily 0908.3451}}]. [Erratum:
  JHEP 03, 033 (2014)].

\bibitem{Ibe:2006de}
M.~Ibe, T.~Moroi and T.~T. Yanagida, \textit{{Possible Signals of Wino LSP at
  the Large Hadron Collider}},
  \href{https://doi.org/10.1016/j.physletb.2006.11.061}{\textit{Phys. Lett. B}
  {\bfseries 644} (2007) 355--360},
  [\href{https://arxiv.org/abs/hep-ph/0610277}{{\ttfamily hep-ph/0610277}}].

\bibitem{Strumia:2006db}
A.~Strumia and F.~Vissani, \textit{{Neutrino masses and mixings and...}},
  \href{https://arxiv.org/abs/hep-ph/0606054}{{\ttfamily hep-ph/0606054}}.

\bibitem{Li:2009mw}
T.~Li and X.-G. He, \textit{{Neutrino Masses and Heavy Triplet Leptons at the
  LHC: Testability of Type III Seesaw}},
  \href{https://doi.org/10.1103/PhysRevD.80.093003}{\textit{Phys. Rev. D}
  {\bfseries 80} (2009) 093003},
  [\href{https://arxiv.org/abs/0907.4193}{{\ttfamily 0907.4193}}].

\bibitem{Fuks:2020zbm}
B.~Fuks, J.~Neundorf, K.~Peters, R.~Ruiz and M.~Saimpert, \textit{{Probing the
  Weinberg operator at colliders}},
  \href{https://doi.org/10.1103/PhysRevD.103.115014}{\textit{Phys. Rev. D}
  {\bfseries 103} (2021) 115014},
  [\href{https://arxiv.org/abs/2012.09882}{{\ttfamily 2012.09882}}].

\bibitem{Atre:2005eb}
A.~Atre, V.~Barger and T.~Han, \textit{{Upper bounds on lepton-number violating
  processes}}, \href{https://doi.org/10.1103/PhysRevD.71.113014}{\textit{Phys.
  Rev. D} {\bfseries 71} (2005) 113014},
  [\href{https://arxiv.org/abs/hep-ph/0502163}{{\ttfamily hep-ph/0502163}}].

\bibitem{Heeck:2018nzc}
J.~Heeck, M.~Lindner, W.~Rodejohann and S.~Vogl, \textit{{Non-Standard Neutrino
  Interactions and Neutral Gauge Bosons}},
  \href{https://doi.org/10.21468/SciPostPhys.6.3.038}{\textit{SciPost Phys.}
  {\bfseries 6} (2019) 038},
  [\href{https://arxiv.org/abs/1812.04067}{{\ttfamily 1812.04067}}].

\bibitem{Akimov:2017ade}
{\scshape COHERENT} collaboration, D.~Akimov et~al., \textit{{Observation of
  Coherent Elastic Neutrino-Nucleus Scattering}},
  \href{https://doi.org/10.1126/science.aao0990}{\textit{Science} {\bfseries
  357} (2017) 1123--1126}, [\href{https://arxiv.org/abs/1708.01294}{{\ttfamily
  1708.01294}}].

\bibitem{Akimov:2018ghi}
{\scshape COHERENT} collaboration, D.~Akimov et~al., \textit{{COHERENT 2018 at
  the Spallation Neutron Source}},
  \href{https://arxiv.org/abs/1803.09183}{{\ttfamily 1803.09183}}.

\bibitem{Esteban:2018ppq}
I.~Esteban, M.~C. Gonzalez-Garcia, M.~Maltoni, I.~Martinez-Soler and
  J.~Salvado, \textit{{Updated constraints on non-standard interactions from
  global analysis of oscillation data}},
  \href{https://doi.org/10.1007/JHEP08(2018)180}{\textit{JHEP} {\bfseries 08}
  (2018) 180}, [\href{https://arxiv.org/abs/1805.04530}{{\ttfamily
  1805.04530}}]. [Addendum: JHEP 12, 152 (2020)].

\bibitem{Sirunyan:2018nnz}
{\scshape CMS} collaboration, A.~M. Sirunyan et~al., \textit{{Search for an
  $L_{\mu}-L_{\tau}$ gauge boson using Z$\to4\mu$ events in proton-proton
  collisions at $\sqrt{s} =$ 13 TeV}},
  \href{https://doi.org/10.1016/j.physletb.2019.01.072}{\textit{Phys. Lett. B}
  {\bfseries 792} (2019) 345--368},
  [\href{https://arxiv.org/abs/1808.03684}{{\ttfamily 1808.03684}}].

\bibitem{TheBABAR:2016rlg}
{\scshape BaBar} collaboration, J.~P. Lees et~al., \textit{{Search for a muonic
  dark force at BABAR}},
  \href{https://doi.org/10.1103/PhysRevD.94.011102}{\textit{Phys. Rev. D}
  {\bfseries 94} (2016) 011102},
  [\href{https://arxiv.org/abs/1606.03501}{{\ttfamily 1606.03501}}].

\bibitem{Aaij:2019bvg}
{\scshape LHCb} collaboration, R.~Aaij et~al., \textit{{Search for
  $A'\to\mu^+\mu^-$ Decays}},
  \href{https://doi.org/10.1103/PhysRevLett.124.041801}{\textit{Phys. Rev.
  Lett.} {\bfseries 124} (2020) 041801},
  [\href{https://arxiv.org/abs/1910.06926}{{\ttfamily 1910.06926}}].

\bibitem{Aad:2019fac}
{\scshape ATLAS} collaboration, G.~Aad et~al., \textit{{Search for high-mass
  dilepton resonances using 139 fb$^{-1}$ of $pp$ collision data collected at
  $\sqrt{s}=$13 TeV with the ATLAS detector}},
  \href{https://doi.org/10.1016/j.physletb.2019.07.016}{\textit{Phys. Lett. B}
  {\bfseries 796} (2019) 68--87},
  [\href{https://arxiv.org/abs/1903.06248}{{\ttfamily 1903.06248}}].

\bibitem{Aaboud:2017sjh}
{\scshape ATLAS} collaboration, M.~Aaboud et~al., \textit{{Search for
  additional heavy neutral Higgs and gauge bosons in the ditau final state
  produced in 36 fb$^{-1}$ of pp collisions at $ \sqrt{s}=13 $ TeV with the
  ATLAS detector}},
  \href{https://doi.org/10.1007/JHEP01(2018)055}{\textit{JHEP} {\bfseries 01}
  (2018) 055}, [\href{https://arxiv.org/abs/1709.07242}{{\ttfamily
  1709.07242}}].

\bibitem{Mishra:1991bv}
{\scshape CCFR} collaboration, S.~R. Mishra et~al., \textit{{Neutrino tridents
  and W Z interference}},
  \href{https://doi.org/10.1103/PhysRevLett.66.3117}{\textit{Phys. Rev. Lett.}
  {\bfseries 66} (1991) 3117--3120}.

\bibitem{Altmannshofer:2014pba}
W.~Altmannshofer, S.~Gori, M.~Pospelov and I.~Yavin, \textit{{Neutrino Trident
  Production: A Powerful Probe of New Physics with Neutrino Beams}},
  \href{https://doi.org/10.1103/PhysRevLett.113.091801}{\textit{Phys. Rev.
  Lett.} {\bfseries 113} (2014) 091801},
  [\href{https://arxiv.org/abs/1406.2332}{{\ttfamily 1406.2332}}].

\bibitem{Jegerlehner:2009ry}
F.~Jegerlehner and A.~Nyffeler, \textit{{The Muon g-2}},
  \href{https://doi.org/10.1016/j.physrep.2009.04.003}{\textit{Phys. Rept.}
  {\bfseries 477} (2009) 1--110},
  [\href{https://arxiv.org/abs/0902.3360}{{\ttfamily 0902.3360}}].

\bibitem{delAguila:2008ir}
F.~del Aguila, S.~Bar-Shalom, A.~Soni and J.~Wudka, \textit{{Heavy Majorana
  Neutrinos in the Effective Lagrangian Description: Application to Hadron
  Colliders}},
  \href{https://doi.org/10.1016/j.physletb.2008.11.031}{\textit{Phys. Lett. B}
  {\bfseries 670} (2009) 399--402},
  [\href{https://arxiv.org/abs/0806.0876}{{\ttfamily 0806.0876}}].

\bibitem{Bhattacharya:2015vja}
S.~Bhattacharya and J.~Wudka, \textit{{Dimension-seven operators in the
  standard model with right handed neutrinos}},
  \href{https://doi.org/10.1103/PhysRevD.94.055022}{\textit{Phys. Rev. D}
  {\bfseries 94} (2016) 055022},
  [\href{https://arxiv.org/abs/1505.05264}{{\ttfamily 1505.05264}}]. [Erratum:
  Phys.Rev.D 95, 039904 (2017)].

\bibitem{Liao:2016qyd}
Y.~Liao and X.-D. Ma, \textit{{Operators up to Dimension Seven in Standard
  Model Effective Field Theory Extended with Sterile Neutrinos}},
  \href{https://doi.org/10.1103/PhysRevD.96.015012}{\textit{Phys. Rev. D}
  {\bfseries 96} (2017) 015012},
  [\href{https://arxiv.org/abs/1612.04527}{{\ttfamily 1612.04527}}].

\bibitem{Datta:2020ocb}
A.~Datta, J.~Kumar, H.~Liu and D.~Marfatia, \textit{{Anomalous dimensions from
  gauge couplings in SMEFT with right-handed neutrinos}},
  \href{https://doi.org/10.1007/JHEP02(2021)015}{\textit{JHEP} {\bfseries 02}
  (2021) 015}, [\href{https://arxiv.org/abs/2010.12109}{{\ttfamily
  2010.12109}}].

\bibitem{Chala:2020pbn}
M.~Chala and A.~Titov, \textit{{One-loop running of dimension-six
  Higgs-neutrino operators and implications of a large neutrino dipole
  moment}}, \href{https://doi.org/10.1007/JHEP09(2020)188}{\textit{JHEP}
  {\bfseries 09} (2020) 188},
  [\href{https://arxiv.org/abs/2006.14596}{{\ttfamily 2006.14596}}].

\bibitem{Datta:2021akg}
A.~Datta, J.~Kumar, H.~Liu and D.~Marfatia, \textit{{Anomalous dimensions from
  Yukawa couplings in SMNEFT: four-fermion operators}},
  \href{https://doi.org/10.1007/JHEP05(2021)037}{\textit{JHEP} {\bfseries 05}
  (2021) 037}, [\href{https://arxiv.org/abs/2103.04441}{{\ttfamily
  2103.04441}}].

\bibitem{CMS:2020xwi}
{\scshape CMS} collaboration, A.~M. Sirunyan et~al., \textit{{Evidence for
  Higgs boson decay to a pair of muons}},
  \href{https://doi.org/10.1007/JHEP01(2021)148}{\textit{JHEP} {\bfseries 01}
  (2021) 148}, [\href{https://arxiv.org/abs/2009.04363}{{\ttfamily
  2009.04363}}].

\bibitem{ATLAS:2020fzp}
{\scshape ATLAS} collaboration, G.~Aad et~al., \textit{{A search for the dimuon
  decay of the Standard Model Higgs boson with the ATLAS detector}},
  \href{https://doi.org/10.1016/j.physletb.2020.135980}{\textit{Phys. Lett. B}
  {\bfseries 812} (2021) 135980},
  [\href{https://arxiv.org/abs/2007.07830}{{\ttfamily 2007.07830}}].

\bibitem{Cepeda:2019klc}
M.~Cepeda et~al., \textit{{Report from Working Group 2}: {Higgs Physics at the
  HL-LHC and HE-LHC}},
  \href{https://doi.org/10.23731/CYRM-2019-007.221}{\textit{CERN Yellow Rep.
  Monogr.} {\bfseries 7} (2019) 221--584},
  [\href{https://arxiv.org/abs/1902.00134}{{\ttfamily 1902.00134}}].

\bibitem{Parker:2018vye}
R.~H. Parker, C.~Yu, W.~Zhong, B.~Estey and H.~M\"uller, \textit{{Measurement
  of the fine-structure constant as a test of the Standard Model}},
  \href{https://doi.org/10.1126/science.aap7706}{\textit{Science} {\bfseries
  360} (2018) 191}, [\href{https://arxiv.org/abs/1812.04130}{{\ttfamily
  1812.04130}}].

\bibitem{Morel:2020dww}
L.~Morel, Z.~Yao, P.~Clad\'e and S.~Guellati-Kh\'elifa, \textit{{Determination
  of the fine-structure constant with an accuracy of 81 parts per trillion}},
  \href{https://doi.org/10.1038/s41586-020-2964-7}{\textit{Nature} {\bfseries
  588} (2020) 61--65}.

\bibitem{EXO-200:2019rkq}
{\scshape EXO-200} collaboration, G.~Anton et~al., \textit{{Search for
  Neutrinoless Double-$\beta$ Decay with the Complete EXO-200 Dataset}},
  \href{https://doi.org/10.1103/PhysRevLett.123.161802}{\textit{Phys. Rev.
  Lett.} {\bfseries 123} (2019) 161802},
  [\href{https://arxiv.org/abs/1906.02723}{{\ttfamily 1906.02723}}].

\bibitem{ATLAS:2019old}
{\scshape ATLAS} collaboration, G.~Aad et~al., \textit{{Search for the Higgs
  boson decays $H \to ee$ and $H \to e\mu$ in $pp$ collisions at $\sqrt{s} =
  13$ TeV with the ATLAS detector}},
  \href{https://doi.org/10.1016/j.physletb.2019.135148}{\textit{Phys. Lett. B}
  {\bfseries 801} (2020) 135148},
  [\href{https://arxiv.org/abs/1909.10235}{{\ttfamily 1909.10235}}].

\bibitem{Aparici:2009fh}
A.~Aparici, K.~Kim, A.~Santamaria and J.~Wudka, \textit{{Right-handed neutrino
  magnetic moments}},
  \href{https://doi.org/10.1103/PhysRevD.80.013010}{\textit{Phys. Rev. D}
  {\bfseries 80} (2009) 013010},
  [\href{https://arxiv.org/abs/0904.3244}{{\ttfamily 0904.3244}}].

\bibitem{Duarte:2015iba}
L.~Duarte, J.~Peressutti and O.~A. Sampayo, \textit{{Majorana neutrino decay in
  an Effective Approach}},
  \href{https://doi.org/10.1103/PhysRevD.92.093002}{\textit{Phys. Rev. D}
  {\bfseries 92} (2015) 093002},
  [\href{https://arxiv.org/abs/1508.01588}{{\ttfamily 1508.01588}}].

\bibitem{Cottin:2021lzz}
G.~Cottin, J.~C. Helo, M.~Hirsch, A.~Titov and Z.~S. Wang, \textit{{Heavy
  neutral leptons in effective field theory and the high-luminosity LHC}},
  \href{https://doi.org/10.1007/JHEP09(2021)039}{\textit{JHEP} {\bfseries 09}
  (2021) 039}, [\href{https://arxiv.org/abs/2105.13851}{{\ttfamily
  2105.13851}}].

\bibitem{Beltran:2021hpq}
R.~Beltr\'an, G.~Cottin, J.~C. Helo, M.~Hirsch, A.~Titov and Z.~S. Wang,
  \textit{{Long-lived heavy neutral leptons at the LHC: four-fermion
  single-N$_{R}$ operators}},
  \href{https://doi.org/10.1007/JHEP01(2022)044}{\textit{JHEP} {\bfseries 01}
  (2022) 044}, [\href{https://arxiv.org/abs/2110.15096}{{\ttfamily
  2110.15096}}].

\bibitem{Duarte:2016miz}
L.~Duarte, I.~Romero, J.~Peressutti and O.~A. Sampayo, \textit{{Effective
  Majorana neutrino decay}},
  \href{https://doi.org/10.1140/epjc/s10052-016-4301-8}{\textit{Eur. Phys. J.
  C} {\bfseries 76} (2016) 453},
  [\href{https://arxiv.org/abs/1603.08052}{{\ttfamily 1603.08052}}].

\bibitem{Duarte:2016caz}
L.~Duarte, J.~Peressutti and O.~A. Sampayo, \textit{{Not-that-heavy Majorana
  neutrino signals at the LHC}},
  \href{https://doi.org/10.1088/1361-6471/aa99f5}{\textit{J. Phys. G}
  {\bfseries 45} (2018) 025001},
  [\href{https://arxiv.org/abs/1610.03894}{{\ttfamily 1610.03894}}].

\bibitem{Duarte:2018kiv}
L.~Duarte, G.~Zapata and O.~A. Sampayo, \textit{{Final taus and initial state
  polarization signatures from effective interactions of Majorana neutrinos at
  future $e^{+}e^{-}$ colliders}},
  \href{https://doi.org/10.1140/epjc/s10052-019-6734-3}{\textit{Eur. Phys. J.
  C} {\bfseries 79} (2019) 240},
  [\href{https://arxiv.org/abs/1812.01154}{{\ttfamily 1812.01154}}].

\bibitem{Zapata:2022qwo}
G.~Zapata, T.~Urruzola, O.~A. Sampayo and L.~Duarte, \textit{{Lepton collider
  probes for Majorana neutrino effective interactions}},
  \href{https://arxiv.org/abs/2201.02480}{{\ttfamily 2201.02480}}.

\end{thebibliography}\endgroup

\end{document}